\documentclass[aps,prx,amsmath,amssymb,twocolumn,superscriptaddress]{revtex4-2}

\usepackage{graphicx}
\usepackage{bm}
\usepackage{hyperref}
\usepackage[capitalise]{cleveref}

\usepackage[caption=false,font=normalfont]{subfig}
\usepackage{floatrow}

\usepackage{tikz}
\usetikzlibrary{positioning, shapes.geometric}

\usepackage{microtype}

\usepackage{glossaries}
\glsdisablehyper
\newacronym{mps}{MPS}{matrix product states}
\newacronym{mpo}{MPO}{matrix product operator}
\newacronym{ed}{ED}{exact diagonalization}
\newacronym{tn}{TN}{tensor network}
\newacronym{eop}{EOP}{entanglement of purification}
\newacronym{tdvp}{TDVP}{time-dependent variational principle}
\newacronym{tebd}{TEBD}{time-evolving block decimation}
\newacronym{metts}{METTS}{minimally entangled typical thermal states}

\newcommand\Hs{\mathcal{H}}

\DeclareMathOperator{\trace}{tr}
\DeclareMathOperator\vect{vec}
\newcommand{\dd}{\mathrm{d}}

\newcommand\bra[1]{\langle #1 \vert}
\newcommand\ket[1]{\vert #1 \rangle}

\newcommand\ketbra[1]{\ket{#1}\bra{#1}}
\newcommand\set[1]{\{\,#1\,\}}
\newcommand\norm[1]{\|#1\|}
\newcommand\abs[1]{|#1|}
\newcommand\order[1]{\mathcal{O}(#1)}
\newcommand\idmat{\mathbb{I}}
\newcommand\Complx{\mathbb{C}}
\newcommand\Reals{\mathbb{R}}

\renewcommand\vec\boldsymbol

\newcommand\Renyi{Rényi }

\begin{document}

\title{Geometrically Taming Dynamical Entanglement Growth in Purified Quantum States}


\newcommand\tudresdenaffil{Institute of Theoretical Physics, Technische Universit\"at Dresden}
\newcommand\ctqmataffil{W\"urzburg-Dresden Cluster of Excellence ct.qmat, 01062 Dresden, Germany}
\author{Tim Pokart}
\email{tim.pokart@tu-dresden.de}
\affiliation{\tudresdenaffil}
\author{Carl Lehmann}
\affiliation{\tudresdenaffil}
\affiliation{\ctqmataffil}
\author{Jan Carl Budich}
\email{jan.budich@tu-dresden.de}
\affiliation{\tudresdenaffil}
\affiliation{\ctqmataffil}
\date{\today}

\begin{abstract}
    Entanglement properties of purified quantum states are of key interest for two reasons. First, in quantum information theory, minimally entangled purified states define the Entanglement of Purification as a fundamental measure for the complexity of the corresponding physical mixed state. Second, dynamical entanglement growth in purified states represents the main bottleneck for calculating dynamical physical properties on classical computers in the framework of tensor network states. Here, we demonstrate how geometric methods including parallel transport may be harnessed to reduce such dynamical entanglement growth, and to obtain a general prescription for maintaining (locally) optimal entanglement entropy when time-evolving a purified state. Adapting and extending by higher order skew corrections the notion of Uhlmann geometric phases, we reveal the relation between dynamical entanglement growth and the geometry of the Hilbert-Schmidt bundle as the mathematical foundation of purified states. With benchmarks on a non-integrable spin chain model, we compare the computational performance of matrix product state algorithms based on our present geometric disentangling method to previous approaches for taming entanglement growth in purified states. Our findings provide numerical evidence that geometric disentanglers are a powerful approach, superior in various aspects to known methods for disentangling purified states in a range of physically relevant computational scenarios. To exclude the effect of algorithmic imperfections, we also provide a numerically exact analysis for systems of moderate size.
\end{abstract}

\maketitle

\section{Introduction}
Disentangling different forms of correlations and complexity in mixed quantum states~\cite{PhysRevA.54.3824,horodecki2001entanglement,wilde_2013} represents a salient challenge~\cite{preskill2012quantum,PhysRevLett.125.200501,morvan2023phase,Kim2023} in quantum science, both at a fundamental quantum information level and at a practical computational one.
In the context of many-body physics, thermal equilibrium states exhibiting a subtle interplay of both classical and genuine quantum correlations are ubiquitous examples of mixed states~\cite{amico2008entanglement,doi:10.1073/pnas.1703516114}.
A versatile platform for unraveling mixed states is provided by the notion of purification~\cite{PhysRevLett.93.207204,nielsen_chuang_2010}, i.e. their representation as pure states $\ket{\psi}$ in a total Hilbert space $\Hs = \Hs_S \otimes \Hs_A$, consisting of the (physical) system state space $\Hs_S$ extended by an auxiliary space $\Hs_A$.
Besides their conceptual importance for entanglement measures such as \gls{eop}~\cite{Terhal_2002}, purified states are also an integral part of the computational toolbox for simulating quantum dynamics on a classical computer using tensor network states~\cite{Schollw_ck_2011}.

\begin{figure}[htp!]
    \centering
    \sidesubfloat[]{
        \begin{minipage}{0.9\textwidth}
            \includegraphics{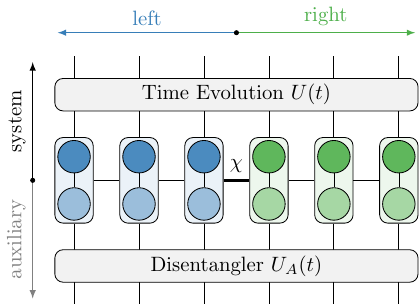}
        \end{minipage}
        \label{fig:mps}
    }
    \\[5pt]
    \sidesubfloat[]{
        \begin{minipage}{0.9\textwidth}
            \includegraphics{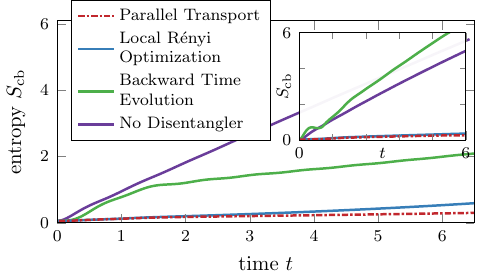}
        \end{minipage}
        \label{fig:plots-tfi_comparison_plot}
    }
    \caption{(a) Illustration of a \gls{mps} purifying a mixed state defined on a lattice of six physical sites (darker) by coupling them to a duplicate of auxiliary sites (lighter). The central bond dimension $\chi$ directly relates to the entanglement entropy $S_{\mathrm{cb}}$ of a spatial bipartition into the left (blue) and right (green) half of the total system. On the system (auxiliary) degrees of freedom a time-evolution (disentangling) operator $U(t)$ ($U_A(t)$) is applied, with drastic effects on entanglement properties. (b) $S_{\mathrm{cb}}(t)$ of the TFI chain (see \cref{eq:model_heisenberg}) for different disentanglers $U_A(t)$ applied after perturbing a thermal state by applying $(\idmat + iX)_{\mathrm{c}}$ at a central site (cf.\ \cref{eq:model_heisenberg}). Specifically, the parallel-transport disentangler introduced here is compared to several established methods. Parameters are $N = 140 = N_S + N_A, \beta = 0.2$, $J_x = -1.0$, $J_z = -0.1$ and $h=-1.0$. 
    The inset shows data obtained for the double field Ising chain (cf. \cref{eq:tranverse_parallel_ising}, $N = 140$, $\beta=0.1$, $J = 1.0$, $g = -1.05$, $h=0.5$) with the full parallel transport being approximated by the $(\beta=0)$ approximation fit for use in gate-based \gls{tn} methods.}
\end{figure}

Specifically, given a purification $\ket{\psi} \in \Hs$, the corresponding physical mixed state $\rho$ is obtained by tracing out the auxiliary degrees of freedom, i.e.  $\rho = \trace_A \ketbra{\psi}$.
This naturally introduces a giant gauge degree of freedom~\footnote{With $D = \textrm{dim}(\Hs_A)$, the gauge group $\textrm{U}(D)$ has $D^2$ real parameters, where $D$ is typically exponentially large in the physical system size. For dynamical problems, $U_A(t)$ may be seen as a local gauge transformation in time.}, since any unitary transformation $U_A$ within the auxiliary space $\Hs_A$ will leave $\rho$ as the result of the partial trace unaffected.
However, for a different partition $\Hs = \Hs_L \otimes \Hs_R$ involving the division of the physical system into left and right spatial regions (see \cref{fig:mps}), the choice of $U_A$ may have a drastic effect on the entanglement properties of $\ket{\psi}$.
Yet, the minimal entropy of say $\rho_L = \trace_R \ketbra{\psi}$ obtained by an optimal choice of $U_A$ defines the \gls{eop}~\footnote{The \gls{eop} in Ref.~\cite{Terhal_2002} is defined with respect to a minimization of both the $U_A$ and the extension $\Hs_A$ itself, while in our case $\Hs_A$ is fixed. This is however only a technicality since an expansion of the auxiliary Hilbert space does not come with increased \gls{eop} as long as $\Hs_A$ is sufficiently large~\cite{bengtsson_zyczkowski_2006}.}, a physical property of $\rho$ encoding its gauge-independent correlations.
Below, we explore the potential of concepts from differential geometry such as geometric phases for solving the outstanding problem of finding an optimal gauge $U_A$ to mitigate entanglement growth in quantum dynamics~\cite{Hastings_2007}.

While our geometric approach generally applies to time evolution of purifications, we provide numerical data on two important computational use-cases where it is immediately beneficial. First, operator evolution after local (unitary) quenches~\cite{PhysRevB.89.075139} of a system equilibrated at high temperatures:
\begin{align}
    \langle O(t) \rangle_{\beta, Q} = \trace [Q \rho_\beta Q^{\dagger} O(t)],
    \label{eq:thermal_expval}
\end{align}
where a mixed quantum state naturally occurs in the form of a thermal state operator $\rho_\beta = Z^{-1} \exp(-\beta H)$ of a system governed by the Hamiltonian $H$, with $Z = \trace\exp(-\beta H)$ the partition function at inverse temperature $\beta = 1/T$. 
In \cref{eq:thermal_expval}, $t$ denotes time such that $O(t)=U^{\dagger}(t) O U(t)$ is a Heisenberg picture operator acting on $\Hs_S$. 
When purifying \cref{eq:thermal_expval}, a typical choice is to model the auxiliary system as a duplicate of the physical system, i.e.  $\Hs_S \cong \Hs_A$ (see \cref{fig:mps}).
Second, we discuss the computation of generic retarded response functions
\begin{align}
    C^R(t) = -i\theta(t)\langle [X(t),Y] \rangle_\beta =  -i\theta(t)\trace( [X(t), Y] \rho_\beta),
    \label{eq:retarded_response_function}
\end{align}
at high temperatures with $[X(t), Y]$ the commutator between the operators $X(t)=U^{\dagger}(t) X U(t)$ and $Y=Y(0)$.

Impressive computational progress on evaluating dynamical correlation functions has been made with the advent of \gls{tn} methods such as typicality-based methods~\cite{PhysRevB.106.094409,PhysRevResearch.3.L022015}, \gls{mpo} operator~\cite{PhysRevB.106.115117} and, most relevant for this work, \gls{mps} approaches~\cite{PhysRevLett.69.2863,PhysRevLett.90.227902,PhysRevLett.93.207204,PhysRevLett.107.070601,PAECKEL2019167998,PhysRevLett.132.100402}.
In the latter, the entanglement of the purified state $\ket{\psi}$ represents the main bottleneck~\cite{Karrasch_2013,Barthel2013} to the quality of numerical data on physical observables.
This is because the computational effort of \gls{mps} simulations, as measured by the required size $\chi$ of the tensors (bond dimension), grows exponentially with the entanglement entropy $S_L(t) = -\trace[\rho_L(t) \log \rho_L(t)]$ (see \cref{fig:mps}), i.e. $\chi(t) \geq \exp(S_L(t))$~\cite{Schuch_2008}.
While $S_L$ is bounded from below by the \gls{eop} relating to the inherent complexity of the physical state, practically saturating this bound by a gauge choice $U_A(t)$ is a hard optimization problem the exact solution to which remains elusive.
However, significant progress for mitigating dynamical entanglement growth by means of a gauge (or disentangler) $U_A(t)$ has been reported based on heuristic arguments~\cite{PhysRevLett.108.227206,Karrasch_2013} and numerical optimization of a variational $U_A(t)$ consisting of local gates~\cite{PhysRevB.98.235163}, respectively.

\begin{figure}
    {
        \vbox to 0pt {
                \raggedright
                \textcolor{white}{
                    \subfloatlabel[1][fig:entanglement_landscape:a]
                    \subfloatlabel[2][fig:entanglement_landscape:b]
                    \subfloatlabel[3][fig:entanglement_landscape:c]
                    \subfloatlabel[4][fig:entanglement_landscape:d]
                }
            }
    }
    \includegraphics{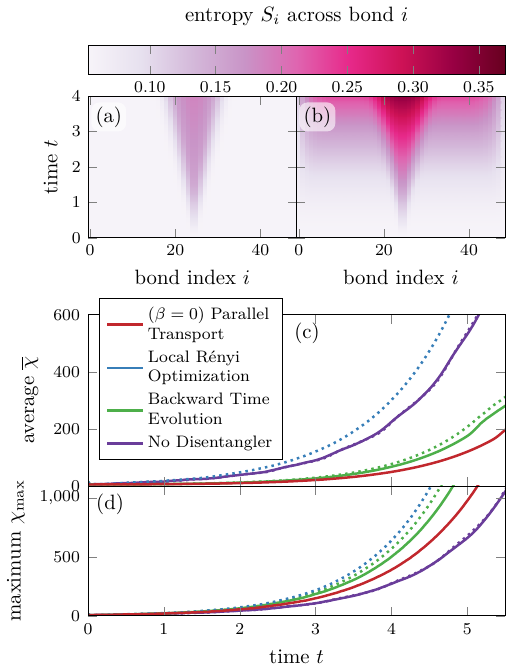}
    \caption{\protect\subref{fig:entanglement_landscape:a} -  \protect\subref{fig:entanglement_landscape:b} Spatio-temporal entanglement landscape of a thermal initial state perturbed by $(\idmat + iX)_{\mathrm{c}}$ at $t=0$ for the TFI model given in \cref{eq:model_heisenberg} with a total system size of $N=100$ sites, comparing the parallel transport disentangler \protect\subref{fig:entanglement_landscape:a}, and the \Renyi disentangler \protect\subref{fig:entanglement_landscape:b}. In panel \protect\subref{fig:entanglement_landscape:c} and \protect\subref{fig:entanglement_landscape:d}, the required maximum and average bond dimension to faithfully represent the purification under time evolution with different disentanglers computed in TeNPy (dotted) and ITensor (solid) are depicted. While the parallel transport disentangler produces states with lower average bond dimension in \protect\subref{fig:entanglement_landscape:c} than the other methods, its maximum bond dimension in \protect\subref{fig:entanglement_landscape:d} is not as small as expected from the significantly decreased von Neumann entropy.}
    \label{fig:entanglement_landscape}
\end{figure}

Here, we report on our results for optimizing the disentangler $U_A(t)$, using as a guiding principle geometric methods including Uhlmann parallel transport of purifications~\cite{UHLMANN1986229,UHLMANN1993253} and the higher order corrections derived below. 
As parallel transport amounts to applying the least change to $\ket{\psi(t)}$ that is compatible with the correct physical trajectory $\rho(t)$, it is quite intuitive that this helps to avoid creating unnecessary correlations in $\ket{\psi(t)}$ as induced by a poor gauge.
More quantitatively, we indeed find that parallel transport of the purified state can substantially reduce entanglement, and is at least competitive with numerically more costly brute force numerical optimization methods in reducing the entanglement growth~\cite{PhysRevB.98.235163} (see \cref{fig:plots-tfi_comparison_plot}).
Moreover, we find that parallel transport can be extended by skew-corrections accounting for differential deviations from a locally optimal entropy $S_L(t)$ to achieve even better results (see \cref{fig:corrected-entropy}), which highlights the far-reaching potential of geometric optimization.
Practically, parallel transport amounts to computing an effective Hamiltonian $G$ that generates the auxiliary time evolution $U_A(t)$.
Interestingly, we find that approximations to $G$ such as its value at infinite temperature can lead to very promising computational results for paradigmatic non-integrable models (see \cref{fig:entanglement_landscape}).
Our general findings are corroborated by numerical benchmark simulations using both \gls{ed} and \gls{mps} methods.

\paragraph*{Outline.} This article is structured as follows:
In \cref{sec:purification} the purification approach to represent mixed states by introducing an auxiliary Hilbert space is briefly reviewed in the differential geometric framework of the Hilbert-Schmidt bundle.
In this context, a parallel transport condition is derived and adapted to the computation of dynamical correlation functions (see \cref{subsec:parallel-transport}). In \cref{subsec:nav-entropy-landscape} these concepts are linked to the dynamical growth of entanglement under time evolution, and higher order corrections to plain parallel transport are derived that allow for maintaining (locally) optimal entanglement.
Thereafter, in \cref{sec:benchmark}, based on two paradigmatic model Hamiltonians of non-integrable spin chains, numerical results in the framework of \gls{ed} (see \cref{subsec:exact_studies}) as well as \gls{mps} (see \cref{subsec:tensor_network_studies}) are provided.
In \cref{subsec:correlation_functions} we discuss applying parallel transport to correlation functions.
Finally, a concluding discussion is presented in \cref{sec:conclusions}.

\section{Geometric Perspective on Entanglement of Purifications}
\label{sec:purification}

\begin{figure}
    \centering
    \includegraphics{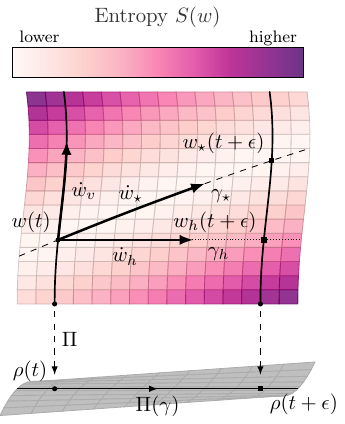}
    \caption{Visualization of the relation between geometry and entanglement landscape in the Hilbert-Schmidt bundle of purifications $w$. In a slice of the bundle that projects to a path $t \mapsto \rho(t)$ of physical mixed states, the tangent vectors $\dot w$ of several distinguished paths $\gamma$ are shown, while the entropy $S(w)$ is quantified by the color code. Specifically, the vertical component $\dot{w}_v$ originating from transport within a given fiber $\Pi^-(\rho(t))$, the horizontal direction $\dot{w}_h$ corresponding to Uhlmann parallel transport, and the entanglement optimal direction $\dot{w}_\star$ following the valley of the entropy are depicted. Importantly, for an initial state that is minimally entangled within its fiber $\Pi^-(\rho(t))$, the optimal route (denoted by $\gamma_*$) is not purely horizontal (cf.~$\gamma_h$), since the entanglement landscape is skewed. Locally however, without further knowledge of the curvature, the horizontal path along $\gamma_h$ (dotted) is the best guess for the direction to move in to leading order.}
    \label{fig:hilbert-schmidt-bundle}
\end{figure}

To purify a (mixed) quantum state described by the density matrix $\rho$ in the \emph{physical} Hilbert space $\Hs_S$, an \emph{auxiliary} Hilbert space $\Hs_A$ is introduced such that a pure state $\ket{\psi} \in \Hs$ in the \emph{total} Hilbert space $\Hs = \Hs_S \otimes \Hs_A$ fulfills
\begin{align}
    \rho = \trace_A \ketbra{\psi}
    \label{eq:partialtrace}
\end{align}
with $\trace_A$ denoting the partial trace of the auxiliary Hilbert space.
This construction admits to an interesting gauge degree of freedom:
given a unitary $U_A$ acting on $\Hs_A$, the transformed purification  $\idmat\otimes U_A \ket{\psi}$ yields the same state $\rho$ as is clear from \cref{eq:partialtrace}.
In \cref{subsec:parallel-transport}, a brief synopsis of how this gauge freedom is used to derive a parallel transport condition \cite{UHLMANN1986229} is presented, singling out the shortest path a purification can traverse in $\Hs$ during a time evolution constrained by the time-dependent physical state $t \mapsto \rho(t)$ in $\Hs_S$.
Then, in \cref{subsec:nav-entropy-landscape}, we relate this parallel transport prescription and higher-order corrections to entanglement growth in quantum dynamics.

For our geometric analysis, we find it helpful to define the auxiliary space as the dual of the physical space, i.e. $\Hs_A = \Hs_S^*$~\footnote{The Schmidt-theorem~\cite{bengtsson_zyczkowski_2006,viennot2018purification} requires that $\dim \Hs_A \geq \dim \Hs_S$ for an arbitrary state to be representable using a purification. Therefore, a choice with $\dim \Hs_A = \dim \Hs_S$ is sufficient.}.
Under this definition, the purification is conveniently represented as a matrix $w \in  \Hs_S \otimes \Hs_S^*$ known as a Hilbert-Schmidt operator.
The Hilbert-Schmidt operator $w$ is equivalent to the conventional purification $\ket{\psi} \in \Hs_S \otimes \Hs_S$ in the sense of the following duality in Schmidt decomposition~\footnote{The set $\set{\ket{\phi_i}}$ refers to a basis of the system Hilbert space $\Hs_S$, while $\set{\ket{\alpha_i}}$ ($\set{\bra{\alpha_i}}$) refers to a basis of the auxiliary Hilbert space $\Hs_A = \Hs_S$ ($\Hs_A = \Hs_S^*$)}:
\begin{align}
    \ket{\psi} = \sum_i \sqrt{\lambda_i} \ket{\phi_i} \ket{\alpha_i} \mapsto w = \sum_i \sqrt{\lambda_i} \ket{\phi_i} \bra{\alpha_i},
\end{align}
which defines a natural isomorphism between the two approaches to purification through flipping the auxiliary basis states from ket to bra type.
Tracing out the auxiliary degrees of freedom to compute the physical density matrix then simply amounts to a matrix multiplication such that \cref{eq:partialtrace} becomes $\rho = w w^\dagger$.
Note that a unitary $U_A$ acting on $\Hs_S^*$ still represents a gauge degree of freedom, since $\tilde{w} = w U_A$ corresponds to the same state $\rho = w w^\dag = \tilde w\tilde w^\dag$.

\subsection{Parallel Transport for Expectation Values}
\label{subsec:parallel-transport}
Turning to differential geometry, the purification may be formalized~\cite{bengtsson_zyczkowski_2006,Budich_2015} as a principal fiber bundle called the Hilbert-Schmidt bundle (see \cref{fig:hilbert-schmidt-bundle}) via the projection
\begin{align}
    \Pi: w \mapsto \rho = w w^\dagger
\end{align}
mapping elements $w$ of the total space of (invertible) Hilbert-Schmidt operators to density matrices $\rho$ of full rank (e.g. thermal states) as elements of the physical base space (see e.g. \cite{Budich_2015} for a more detailed discussion).
The typical fiber (gauge group) consists of the unitary transformations $U_A$ on the auxiliary (dual) space $\Hs_A = \Hs_S^*$.
In addition, the Hilbert-Schmidt bundle has a metric structure as induced by the Hilbert-Schmidt inner product
\begin{align}
    (w, v) = \trace w^\dagger v
    \label{eq:scalar_prod}
\end{align}
of two matrices $w$ and $v$.
Finally, the mapping $\rho \rightarrow \sqrt{\rho}$ uniquely defines a global section on the Hilbert-Schmidt bundle as a consequence of the polar decomposition of $w$, thus implying a topologically trivial global product structure.

Parallel transport and geometric phases in quantum mechanics are best known for pure states resulting in the accumulation of a Berry phase~\cite{doi:10.1098/rspa.1984.0023} when the Hamiltonians parameters are varied adiabatically.
Uhlmann~\cite{UHLMANN1986229,UHLMANN1993253} derived a way to extend the Berry phase formalism to generic density matrices.
Given a path of purifications $\gamma: t \mapsto w(t)$ in the Hilbert-Schmidt bundle, an arbitrary element $\dot{w} \equiv \frac{\dd}{\dd t} w$ of the tangent space at $w$ can be decomposed into a vertical part $\dot{w}_v$ (generating a pure gauge transformation) and a horizontal part $\dot{w}_h$ (defining the direction of parallel transport) with respect to the path $\gamma$.
To construct the vertical vector $\dot{w}_v$~\cite{DITTMANN199293,Budich_2015}, we consider at first a path $\gamma' : t \mapsto w(t) = w_0 U_A(t)$, which stays within a single fiber of the bundle, i.e. amounts to a gauge transformation.
The unitary operation $U_A(t)$ describing the path, is at $t=0$ generated by a Hermitian matrix $g$ such that $U_A(t) = \exp (-i t g)$.
Then, the vertical vectors at $t=0$ are given by
\begin{align}
    \dot{w}_v = \frac{\dd}{\dd t} w_0 e^{-i t g}\rvert_{t=0} = - i w_0  g.
\end{align}
The respective horizontal vectors $\dot{w}_h$ along the unrestricted path $\gamma$ can then be defined via the scalar product \cref{eq:scalar_prod} requiring orthogonality with respect to the vertical components such that
\begin{align}
    0 & = (\dot{w}_v, \dot{w}_h) + (\dot{w}_h, \dot{w}_v) = - i \trace (g (\dot{w}_h^\dagger w - w^\dagger \dot{w}_h)),
\end{align}
irrespective of the hermitian matrix $g$.
Hence, a tangent vector $\dot{w}$ of an arbitrary path is horizontal with respect to a purification $w$ if and only if \cite{UHLMANN1986229}
\begin{align}
    \label{eq:parallel_transport}
    \dot{w}^\dagger w - w^\dagger \dot{w} = 0.
\end{align}
This condition can be formally solved for a general time dependent invertible density matrix $\rho(t) = w(t)w^{\dagger}(t)$.
The resulting unitary part $\mathcal U(t)$ of the amplitude $t \mapsto w(t)=\sqrt{\rho(t)}\mathcal U(t)$ along a parallel transported path is given by~\cite{Viyuela_2015,Budich_2015}
\begin{align}
    \mathcal U(t) = \mathcal{T} e^{-\int_{\gamma}\mathcal{A}}\, U_A(0),
\end{align}
where $\mathcal{T}$ denotes time-ordering and the Uhlmann connection $\mathcal{A}=\mathcal U(\dd \mathcal U)^{\dagger}$ may be obtained as the solution to a Lyapunov equation yielding~\cite{UHLMANN1986229}
\begin{align}
    \mathcal{A} = \int_{0}^{\infty} \dd\tau e^{-\tau \rho(t)} [\sqrt{\rho(t)},\dd\sqrt{\rho(t)}] e^{-\tau \rho(t)}.
\end{align}
The total phase change $H_{\gamma}=\mathcal U(t) \mathcal U^{\dagger}(0)$ is known as the Uhlmann geometric phase (holonomy).

For our purposes, an important special case is to assume a time-evolution $t \mapsto \rho(t)= e^{-iHt}\rho(0)e^{iHt}$ generated by a time-independent physical Hamiltonian $H$.
The time evolution on the auxiliary space then simplifies to $\mathcal{U}(t)= e^{-iHt} e^{-iGt} \mathcal{U}(0)$ with the time-independent auxiliary Hamiltonian $G$ that is of the explicit form
\begin{align}
    G = -2 \int_0^\infty \dd \tau e^{-\rho(0)\tau} \sqrt{\rho(0)} H \sqrt{\rho(0)} e^{-\rho(0)\tau}.
    \label{eq:aux_ham_formal}
\end{align}
The equivalent Lyapunov equation in this case reads as
\begin{align}
    A G + G A + Q = 0
    \label{eq:lyapunov}
\end{align}
with $A = -\rho(0)$ and $Q = -2\sqrt{\rho(0)} H \sqrt{\rho(0)}$.

\paragraph*{Limits of the parallel transport Hamiltonian.}
We now discuss several analytically tractable limits of \cref{eq:aux_ham_formal}.
To this end, we consider the purification $w = Q w_\beta$ with $w_\beta$ the purification of the finite temperature density matrix $\rho_\beta$ and $Q$ the local perturbation in \cref{eq:thermal_expval}.
First, both in the case of no perturbation, i.e. $Q=\idmat$, and in the zero temperature limit $\beta = \infty$ ($T = 0$), we find that $G = -H$, i.e. the program of parallel transport reduces to backward time evolution in the sense of Ref.~\cite{Karrasch_2013}. 
This corroborates the theoretical foundation of the heuristically introduced backward time evolution approach.
Second, the limit of infinite temperature $\beta = 0$ ($T = \infty$) in \cref{eq:aux_ham_formal} for which the auxiliary Hamiltonian assuming a {\em{unitary}} operator $Q$~\footnote{Unitarity of $Q$ up to a constant, i.e. $Q Q^\dagger = c \idmat$ with $c \neq 0$ and $c \in \Complx$ is sufficient.} reads as
\begin{align}
    G = -Q^\dagger H Q.
    \label{eq:g_inf}
\end{align}
This practically amounts to applying the (conjugate transposed) operator on the auxiliary space and time evolving the auxiliary degrees of freedom with the system Hamiltonian.
Remarkably, as shown in \cref{subsec:tensor_network_studies}, the infinite temperature limit in \cref{eq:g_inf} yields a simple and widely applicable approximation even for physical states at finite temperature.

\paragraph*{Adaption for correlation functions.}
While it is straightforward to apply parallel transport to expectation values such as \cref{eq:thermal_expval}, in order
to harness parallel transport for the calculation of the correlation function in \cref{eq:retarded_response_function}, we introduce the generalized correlation function \footnote{To see this, consider the correlator $\langle X(t) Y \rangle$ in the infinite temperature limit of $G$ and 
\begin{align*}
    &\langle X(t) Y \rangle \\
    &\quad= \trace [e^{-\beta/2 H}U^\dagger(t) X U(t) Ye^{-\beta/2 H}] \\
    &\quad= \trace [e^{-\beta/2 H} Y_A^\dagger e^{iHt} Y_A X Y_A^\dagger e^{-iHt}Y_A Ye^{-\beta/2 H}],
\end{align*}
such that the term $e^{-\beta/2 H} Y_A^\dagger e^{iHt}$ on the left-hand side corresponds to the backward time evolution of the operator $Y_A$ acting on the auxiliary system and hence limits the success of any improvement on the right-hand side.
}
\begin{align}
    \hat{C} (\mu, t) = \trace [X(t) \rho_\mu]
    \label{eq:hatCmu}
\end{align}
with $\rho_\mu = (\idmat + \mu Y) \rho (\idmat + \mu Y)^\dagger$ and $\mu \in \Complx$.
Then, the retarded correlation function involving the commutator $[X(t), Y]$  of $X(t)$ and $Y$ can be obtained from 
\begin{align}
    C^-(t) & = \frac{\hat{C}(i\delta, t) - \hat{C}(-i\delta, t)}{2i\delta} \label{eq:corr_fun_retarded}
\end{align}
with $\delta \in \Reals$. While \cref{eq:corr_fun_retarded} holds exactly for any finite $\delta$, its form is reminiscent of a finite difference approximation to a derivative~\footnote{In \cref{app:correlation_functions_with_phase_matching}, we provide a detailed discussion of the differential representation of the equations for computing correlation functions.}.  
Now, since $\rho_\mu$ are still Hermitian matrices of full rank, their purification $w_\mu$ is well defined up to the aforementioned gauge degree of freedom $U_A$, and
\begin{align}
    \hat{C}(\mu, t) = \trace[w_\mu^\dag X(t) w_\mu] = \trace[U_A^\dag(t) w_\mu^\dag X(t) w_\mu U_A(t)]
    \label{eq:hatCmuTraceInvariance}
\end{align}
are gauge invariant functions due to the cyclic invariance of the trace. 
Hence, gauge-optimizing the time-dependent purifications $\tilde w_\mu(t) = U(t) w_\mu U_A(t)$ represents a well-defined problem that can be directly addressed using the geometric toolbox of parallel transport. 

This mechanism can be expanded to include the correlation functions involving anticommutators, which are retrieved from 
\begin{align}
    C^+(t) = \langle \{X(t),Y\} \rangle_\beta = \frac{\hat{C}(\delta, t) - \hat{C}(-\delta, t)}{2\delta}. \label{eq:corr_fun_advanced}
\end{align}
Consequently, arbitrary correlation functions of the form 
\begin{align}
    C(t) = \langle X(t) Y\rangle_\beta
\end{align}
may be assembled in this framework from commutator and anticommutator contributions as 
\begin{align}
    C(t) = \frac{C^-(t) + C^+(t)}{2}.
    \label{eq:correlator_prescription}
\end{align}
However, this formal generality should be taken with a grain of salt: as we demonstrate below, the practical success of geometric disentanglers is largely limited to the case of unitary perturbations applied to $\rho$ (see \cref{eq:corr_fun_retarded}, and \cref{fig:corrected-entropy,fig:corrected-entropy-real} for unitary and non-unitary entanglement growth), which makes their computational application to commutator-based correlation functions such as \cref{eq:thermal_expval} much more viable than to anti-commutator based correlation functions.

\subsection{Navigating the Entanglement Landscape Beyond Plain Parallel Transport}
\label{subsec:nav-entropy-landscape}

In the total space of purifications $w$, one may define an entanglement entropy function $S(w)$ (see \cref{fig:hilbert-schmidt-bundle} for an illustration), e.g. representing the von Neumann entanglement entropy of the left-right decomposition illustrated in \cref{fig:mps}. The subject of interest is then how $S$ changes for small variations of $w$ and in particular, how purifications of minimal entropy within their individual fiber $\Pi^-(\rho)$ evolve along cross-sections  (tantamount to gauge choices $U_A(t)$).

Generally speaking, reaching the \gls{eop} $E_p(\rho)$ as the principle lower bound on the entropy of interest $S(w)$ for any purification $w$ (with $\rho = w w^\dagger$) represents the ultimate goal of any disentangler. $E_p(\rho)$ is bounded by a Fannes-type inequality~\cite{Fannes1973,Nielsen_2000,Audenaert_2007} that reads as~\cite{Terhal_2002}
\begin{align}
    \abs{E_p(\rho) - E_p(\rho + \dd \rho)} \leq A \norm{\dd \rho} - \norm{\dd \rho} \log  \norm{\dd \rho},
    \label{eq:fannes}
\end{align}
with a constant $A$ depending on the specific Hilbert space under investigation and $\norm{\dd \rho}$ the length of the small deviation $\dd \rho$ from $\rho$ as measured by the Bures distance~\cite{Bures1969}.
The Bures distance is defined on the Hilbert-Schmidt bundle as the minimal path length~\cite{Uhlmann1992}
\begin{align}
    \norm{\dd \rho} = \min_\gamma \int_\gamma \sqrt{(\dot{w}, \dot{w})} \;\dd t
    \label{eq:pathlength}
\end{align}
for paths $\gamma$ in the Hilbert-Schmidt bundle from purifications of $\rho$ to $\rho + \dd \rho$.
This length is precisely minimized by $\gamma$ fulfilling the parallel transport condition \cref{eq:parallel_transport}~\cite{Uhlmann1992,bengtsson_zyczkowski_2006}.
Therefore, from \cref{eq:pathlength} alone, it is intuitive that parallel transported changes in the purified state $w$ translate to small changes in the entropy function as bounded by \cref{eq:fannes}.
In this sense, minimizing \cref{eq:pathlength} along a physical path is intuitively favorable for mitigating dynamical entanglement growth.
However, \cref{eq:fannes} does not provide a monotonous one-to-one relation between entanglement growth and change of the purification, and we indeed find that the plain parallel transport prescription \cref{eq:parallel_transport} needs refinement to reveal a comprehensive geometric perspective on the entanglement entropy, as detailed in the following.

An optimal path $\gamma_*: t \mapsto w_*(t)$ of purifications in the Hilbert-Schmidt  bundle is distinguished by starting from a purification $w_*(0)$, which is minimally entangled within the initial fiber $\Pi^-(\rho(0))$ with $\rho(0)=w_*(0) w_*^\dag(0)$  (optimal initial gauge), and is then transported such that the entanglement remains in a minimum within the fibers $\Pi^-(\rho(t))$ at any later point in time. To formalize this situation, we express the entropy $S(t, \{\theta_i\})$ by parameterizing the purification $w(t, \{\theta_i\})$ in terms of time $t$, and the parameters $\theta_i$ of the linear combination of generators $g_i$ describing the position along the fiber, such that $w(t, \{\theta_i\}) = e^{-i H t} w(0) e^{-i\sum_i \theta_i(t) g_i}$.
The initially minimal entanglement entropy then requires a vanishing gradient $\partial_i S|_{w_*} = 0$ in the local parameters at the starting position.
In order for the purification to continue to be locally optimal during the transport, this stationarity needs to persist, i.e. the path $\gamma_*$ needs to evolve along the valley of minimal $S$ in the total space, which requires~
\begin{align}
    \frac{\dd}{\dd t} \partial_i S = \partial_i \partial_t S + (\partial_i \partial_j S) \dot{\theta}_j = 0.
    \label{eq:entropy-linear-system}
\end{align}
\cref{eq:entropy-linear-system} may be seen as the equation of motion governing entanglement-optimized transport of purifications. The solutions $\gamma_*$ to \cref{eq:entropy-linear-system} generically deviate slightly, for long times sometimes significantly, from the horizontally lifted purifications $\gamma_h$, i.e. the solutions to the plain parallel transport condition \cref{eq:parallel_transport} (see \cref{fig:hilbert-schmidt-bundle} for an illustration). We refer to these deviations as {\em{skew corrections}} to parallel transport.

However, due to the rapid (exponential in system size) growth of the number of parameters $\theta_i$, a numerically exact solution to \cref{eq:entropy-linear-system} is generically hard to obtain, except for small systems (see \cref{subsec:exact_studies}). Therefore, since an initially optimal purification implies that the Hessian $\partial_i \partial_j S$ is positive definite such that generic movements along the vertical direction locally increase entanglement, the best direction to move in without further (higher-order) information is the purely horizontal one of parallel transport, which can be efficiently determined as is discussed in \cref{subsec:parallel-transport} and numerically exemplified in \cref{sec:benchmark}.

In summary, the above analysis provides two key insights explaining with theory the success in mitigating dynamical entanglement growth (see \cref{sec:benchmark} for numerical data) of our present geometric approach. First, plain parallel transport can go a long way to mitigate dynamical entanglement growth since its minimal changes to the purified state result in limited changes to the entanglement entropy. Second, maintaining a minimum of entanglement entropy along a physical path amounts to solving a conceptually simple but numerically challenging equation of motion (see \cref{eq:entropy-linear-system}).

A remarkable simplification again occurs in the infinite temperature limit. There, the exact expression for the parallel transport generator $G$ (see \cref{eq:g_inf}) is found to also solve \cref{eq:entropy-linear-system}, i.e. no skew corrections to parallel transport occur. In addition, since the infinite temperature state does not contain any spatial entanglement, the dynamics is guaranteed to start at the global optimum of the entanglement entropy landscape. This favorable stability of \cref{eq:g_inf} provides theoretical intuition for our numerical finding (see \cref{subsec:tensor_network_studies}) that using the infinite temperature value of $G$ even at finite temperature provides a quite powerful and simple approximation.

\section{Harnessing Mixed State Geometry for Computational Methods}
\label{sec:benchmark}

We now study the consequences of the above geometric analysis on the computation of operators after local quenches (see \cref{eq:thermal_expval}) and correlation functions (see \cref{eq:retarded_response_function}) on a classical computer, both at a conceptual level of entanglement dynamics and regarding practical implications for \gls{tn} method development, where entanglement growth represents the main limitation to simulating time-evolution of quantum many-body systems. In the following, we compare our geometric disentangling method to two major previous approaches:  First, the heuristic idea of applying the physical time-evolution backward on the auxiliary system as a disentangler~\cite{PhysRevLett.108.227206,Karrasch_2013,KENNES201637}, which is both popular and widely used due to its simplicity and effectiveness in significantly mitigating entanglement ~\cite{Barthel2013,Paeckel2019,Rausch2020,Peters2023}. Second, the variational approach of approximating the global (on the auxiliary system) gauge degree of freedom $U_A(t)$ as a temporal series of local gates coupling two neighboring sites, and determining the parameters of those gates by solving an optimization problem for the (second) \Renyi entropy at every time-step \cite{PhysRevB.98.235163}. While computationally more costly, this second approach in principle has the potential to further reduce the entanglement entropy beyond the backward time-evolution method for certain scenarios.

\paragraph*{Model for local quenches.}

As a notoriously difficult model system for quench dynamics, we study the double field Ising Hamiltonian:
\begin{align}
    H = J \sum_{\langle i,j\rangle} Z_i Z_j + \sum_i (h Z_i + g X_i)
    \label{eq:tranverse_parallel_ising}
\end{align} 
with hopping $J=1.0$ as well as field strengths $h = 0.5$ and $g = -1.05$ in the parallel and transverse direction, respectively, chosen to be far from integrability \cite{PhysRevLett.106.050405,PhysRevB.106.115117}. Here $X_i, Y_i$ and $Z_i$ represent the three components of the spin operator at site $i$ with $\langle i, j \rangle$ denoting nearest neighbors.
The chain of $N_S$ spin-$1/2$ sites arranged in a one-dimensional lattice  (forming the system $\Hs_S$) is equilibrated at inverse temperature $\beta = 0.1$, which is created by purifying the identity matrix on $\Hs_S$ (infinite temperature state) in a local singlet configuration that maximally entangles every physical site with its auxiliary duplicate, and then applying imaginary time-evolution~\cite{Schollw_ck_2011} to obtain the purified state $\ket{\psi}$ at finite temperature, now living in a total Hilbert space with $N = N_S + N_A = 2N_S$ sites. This equilibrium state is then perturbed by flipping the $K$ spins closest to the center, i.e. $Q = X_{L/2-K/2} \cdots  X_{L/2+K/2}$; here $K=3$ is chosen. For assessing the anticipated computational complexity, a primary quantity of interest is the von Neumann entanglement entropy $S_i$ across the bond $i$ connecting the sites $i$ and $i + 1$ in a left-right bipartition (see \cref{fig:mps}), defined as $S_i = -\trace [\rho_i \log \rho_i]$ with $\rho_i = \trace_{j > i} \ket{\psi}\bra{\psi}$ being the reduced density matrix after tracing over sites $j>i$ on both system and auxiliary. 
The main focus is then on the half-system bipartition with the central-bond entropy $S_{\mathrm{cb}}$ as the generic worst-case scenario expected to exhibit the largest entanglement entropy, and thus requiring the largest bond dimension $\chi$ in \gls{mps} simulations. 
We note that there are more detailed representability indicators~\cite{Verstraete_2006} provided by \Renyi entropies across the respective bonds, and an in-depth discussion regarding the dependence of the entropy on the \Renyi index will be presented further below. Yet, as a single aggregate figure for assessing entanglement growth, the von Neumann entropy with its direct relation to the Shannon entropy of probability distributions remains appropriate and is most familiar in literature.

\paragraph*{Model for correlation functions.}
To obtain a quantitative comparison of our geometric construction to previous work in the context of a concrete microscopic model system, we focus on the non-integrable extension of the transverse field Ising model studied in Ref.~\cite{PhysRevB.98.235163} as a workhorse Hamiltonian of $N_S$ spin-$1/2$ sites on one-dimensional lattice:
\begin{align}
    H = \sum_{\langle i, j \rangle} (J_x X_i X_j + J_z Z_i Z_j)  + h \sum_i Z_i,
    \label{eq:model_heisenberg}
\end{align}
subject to open boundary conditions. 
Unless explicitly stated otherwise, we choose $J_x = -1.0$,  $J_z = -0.1$ and $h = -1.0$ for the coupling parameters, and consider for the average in \cref{eq:thermal_expval} a thermal state at inverse temperature $\beta = 0.2$. This state is then perturbed by $(\idmat + i X)_c$ as a benchmark operator acting on the central site, representing a typical operator involved in computing retarded spin-spin correlation functions using the scheme introduced in \cref{subsec:parallel-transport}. All other aspects of the simulations on correlation functions follow the same principles as those on the local quench setup described above under \cref{eq:tranverse_parallel_ising}.

\paragraph*{Outline of computational benchmark.}
In the remainder of this section, we will first present exact numerical data in the framework of \gls{ed}~\cite{Julia-2017,Optim.jl-2018,harris2020array,2020SciPy-NMeth} in \cref{subsec:exact_studies} to assess at a fully microscopic level the potential of our geometric disentangling approach, independent of any approximations that may be made for practical computational reasons when using \gls{tn} methods. Afterward, in \cref{subsec:tensor_network_studies}, we generally outline and concretely benchmark for exemplary settings how to harness geometric disentanglers for faster \gls{mps} calculations. Finally, in \cref{subsec:correlation_functions} we briefly discuss applying the parallel transport scheme to correlation function that also involve anti-commutators of operators.

\subsection{Exact Diagonalization Analysis}
\label{subsec:exact_studies}

\begin{figure}
    \centering
    \includegraphics{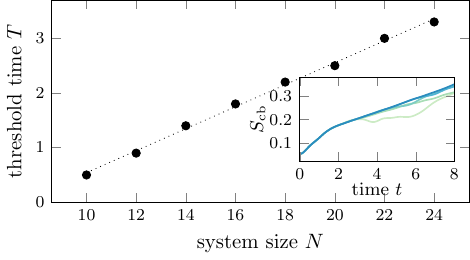}
    \caption{The threshold passing times $T$ as a function of system size $N$ of the TFI chain (see \cref{eq:model_heisenberg}) at which $S_{\mathrm{cb}}$ for plain parallel begins to differ from the curve at $N=26$ by at least $\epsilon_{\mathrm{thr}} = 10^{-6}$.  The observed linear increase with system size suggests that the deviation between the respective curves is mainly a finite-size effect resulting from excitations that travel from the system center to the open boundaries. The inset shows the time dependent central bond entropy $S_{\mathrm{cb}}$ for system sizes $N=10,14,18,22,26$ with increased blueness corresponding to increased system-size $N$.}
    \label{fig:finite_size_effects}
\end{figure}

In order to numerically carry out the parallel transport governed by the auxiliary Hamiltonian $G$, we may rewrite \cref{eq:aux_ham_formal} in terms of the initial purification $w$ as
\begin{align}
    \tilde{G} = -2 \int_0^\infty \dd \tau e^{-w^\dagger w\, \tau} w^\dagger H w e^{-w^\dagger w\, \tau},
    \label{eq:auxham_w}
\end{align}
so as to avoid the polar decomposition required to compute $\sqrt{\rho(0)}$. The explicit evaluation of the integral in \cref{eq:auxham_w} may also be circumvented by obtaining the auxiliary Hamiltonian $\tilde G$ in \cref{eq:auxham_w} directly from the corresponding Lyapunov equation 
\begin{align}
    \tilde A \, \tilde G + \tilde G \, \tilde A + \tilde Q = 0,
    \label{eq:LyapunovNumerical}
\end{align}
where $\tilde A = - w^\dag w,~ \tilde Q = - 2 w^\dag H w$. 
Parallel transport is then described by the time-evolution $t \mapsto w(t) =  e^{-i H t} w e^{-i \tilde{G} t}$, now acting directly on the initial purification $w$.
The MPS data for the \Renyi disentangler \cite{PhysRevB.98.235163} to which we compare our geometric approach is obtained with the help of the tensor network package TeNPy (0.10.0)~\cite{tenpy}. Convergence of the different results is checked by monitoring the Fr\"obenius norm difference of the physical density matrices $\rho(t)$ so as to make sure that only different gauge transformations are compared while maintaining the exact physical time-evolution to high precision in all calculations.

\paragraph*{Finite size effects. }
We begin by confirming that the entropy growth observed during parallel transport is well-behaved with increasing system size, i.e. converges towards a common value at a fixed time.
For finite systems, the open boundaries of the system are expected to become visible to excitations spreading from the center at a certain finite velocity $v$~\cite{Lieb1972,PhysRevLett.97.050401}. 
To verify this intuition, we consider the deviation of plain parallel transport for different system sizes quantified by the accumulated $L^2$ norm distance of two lines $\gamma$ and $\sigma$ at time $t$
\begin{align}
    d(\gamma, \sigma, t) = \sqrt{\int_0^t \dd \tau [S_{\mathrm{cb},\gamma}(\tau) - S_{\mathrm{cb},\sigma}(\tau)]^2}.
\end{align}
By comparing the times $T$ after which this distance passes a small threshold, i.e. $d(\gamma, \sigma, T) \geq \epsilon_{\mathrm{thr}}$, for $\gamma$ corresponding to parallel transport at system sizes $10 \leq N \leq 24$ and $\sigma$ at $N=26$, we measure the times up to which the two lines approximately concur.
The linear increase of $T$ with system size indicated by the data shown in \cref{fig:finite_size_effects} is indeed compatible with the physical picture of a light-cone hitting the open boundaries at $v T = N_S/2$. 
This suggests that the parallel transport is well-behaved on the time-scales and system-sizes relevant for numerical simulations of large systems, i.e. until the open boundaries become visible in the data as a significant finite-size effect.

\begin{figure}
    \includegraphics{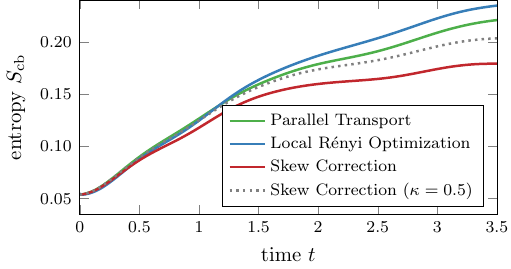}
    \caption{The entropy $S_{\mathrm{cb}}$ across the central bond at $N=12$ sites of the TFI chain (see \cref{eq:model_heisenberg}) for various disentanglers, showing the improvement obtainable by correcting for the skewing of the entanglement landscape. The initial entanglement entropy was optimized using gradient descent for the initial purification to be locally minimal.}
    \label{fig:corrected-entropy}
\end{figure}

\paragraph*{Accounting for skew corrections.} Next, we explicitly compute the skew corrections introduced in \cref{eq:entropy-linear-system} and compare them to those obtainable from the local \Renyi optimization, thus demonstrating that entanglement can in principle be reduced by geometric means even significantly below the values achieved with \Renyi optimization. 
To this end, we parameterize the vertical tangent space along the fibers (see \cref{fig:hilbert-schmidt-bundle}) using a full basis of Hermitian matrices acting on $\Hs_A$ as generators $g_i$, and compute the full Hessian in \cref{eq:entropy-linear-system}.
Solving the linear system using GMRES~\cite{Saad1986}~\footnote{Using a MINRES algorithm in this case is necessary, since the solution to \cref{eq:entropy-linear-system} is not unique. This is due to the fact, that junctions in the entanglement landscape can occur.} then allows us to study the effects of the skew correction. 
Specifically, the numerical data shown in \cref{fig:corrected-entropy} clearly shows an additional reduction of $S_{\mathrm{cb}}$ compared to both plain parallel transport and local \Renyi optimization. 
Despite the skew-corrections success to further mitigate entanglement growth, we note that the exact computation of skew corrections is limited to small systems due to the exponential growth in system size of the number $\mathcal N_g = \mathrm{dim}(\Hs_A)^2$ of generators $g_i$ of the unitary group $U(\mathrm{dim}(\Hs_A))$ of eligible disentangling transformations. 
To assess the dimensional compressibility of the skew correction problem \cref{eq:entropy-linear-system}, we perform an optimal $k$-rank approximation and refer to $\kappa$ as the compression ratio, i.e.\ how many singular values of $\partial_i \partial_j S$ are discarded.
Here, discarding half the singular values ($\kappa = 0.5$, cf.\ \cref{fig:corrected-entropy}) almost completely diminishes the effect of the skew-correction, suggesting that the relevant information is not restricted to a few degrees of freedom, at least not for the small systems under exact consideration.
Still, computing skew corrections is much more viable than brute force global optimization of $U_A(t)$, which would even scale exponentially in the already exponentially big number $\mathcal N_g$. In \cref{subsec:tensor_network_studies}, we will outline how skew corrections may be used at least approximately to go beyond plain parallel transport in \gls{mps} simulations.

\begin{figure}
    \includegraphics{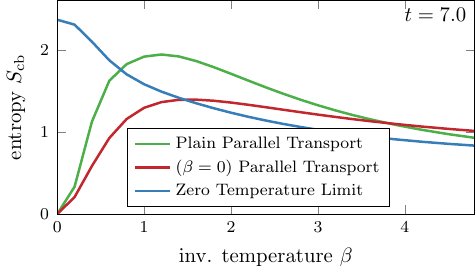}
    \caption{The central-bond entropy $S_{cb}$ at $t=7.0$ compared for the plain parallel transport compared to its infinite temperature ($\beta=0$) and zero temperature  ($\beta = \infty$) limit, showing the large viability regime of the respective approximations. The data is obtained from \gls{ed} calculations at $N=26$ sites of the TFI chain (see \cref{eq:model_heisenberg}).}
    \label{fig:beta_comparison}
\end{figure}

\paragraph*{Viability regime of the temperature limits.}
The two limits to \cref{eq:aux_ham_formal} for zero temperature ($\beta = \infty$) and infinite temperature ($\beta = 0$) are viable at different temperature regimes as shown in \cref{fig:beta_comparison}.
If the respective regimes are not known, the plain parallel transport is an effective guess, while for both low and high temperatures, its limits are better suited to mitigate entanglement growth due to their approximate inclusion of skew corrections.

\subsection{Matrix Product State Benchmark}
\label{subsec:tensor_network_studies}

The variational framework of \gls{tn} methods in general and \gls{mps} as the most widely used \gls{tn} ansatz designed for one-dimensional systems in particular have advanced to an important part of the computational toolbox in recent years. \gls{tn} states provide a numerically unbiased method where the main bottleneck lies in the exponential cost with increasing entanglement entropy of the parametrized states. Mitigating the dynamical growth of entanglement in quantum time evolution is thus an important program for extending the range of classical computers regarding the simulation of quantum many-body systems. In the following, we investigate as to what extent our present geometric disentangler approach can practically contribute to this program by performing \gls{mps} simulations on the model defined in \cref{eq:model_heisenberg} and comparing their performance to previous disentangling approaches.

A pure (or purified) state $\ket{\psi}$ on a lattice comprised of $N$ sites  with open boundary conditions and local basis states $\ket{\sigma_i}$ can be expressed as a \gls{mps}~\cite{perezgarcia2007matrix,Schollw_ck_2011} by decomposing its wavefunction into tensors $A^{\sigma_i}_i$ such that
\begin{align}
    \ket{\psi} = \sum_{\{\sigma_i\}} A^{\sigma_1}_1 \cdots A^{\sigma_L}_L \ket{\sigma_1\ldots\sigma_L}.
\end{align}
For states with sub-extensive entanglement entropy, e.g. for area law entangled states~\cite{RevModPhys.82.277}, this construction allows to reach values of $N$ far beyond the scope of \gls{ed}, because the required matrix size (bond-dimension) of the tensors $A^{\sigma_i}_i$ grows sub-exponentially with system size. Remarkably, for ground states of one-dimensional systems with an energy gap and thus a finite correlation length, the required bond dimension asymptotically does not grow at all with $N$~\cite{Schollw_ck_2005}, thus allowing to solve arbitrary big systems with \gls{mps} methods such as the density matrix renormalization group (DMRG)~\cite{PhysRevLett.69.2863}. Generic operators are included in this framework by decomposing them in a similar manner to obtain a \gls{mpo}, e.g. for the Hilbert-Schmidt operator representation $w$ of the purified state. Importantly, purifications of generic thermal states only contain area law spatial entanglement and can thus be efficiently represented as \gls{mps}~\cite{barthel2017onedimensional}. This property makes them superior to other approaches such as \gls{metts}~\cite{binder2015minimally} for our present context of computing dynamical finite-temperature correlation functions. The main challenge is then to mitigate by means of the disentangler $U_A(t)$ (see \cref{fig:mps}) the dynamical growth of entanglement in the time-evolution after the equilibrium state has been perturbed by a local operator inducing the quasi-particle like propagation of entanglement~\cite{Calabrese_2009}.

\begin{figure}
    \includegraphics{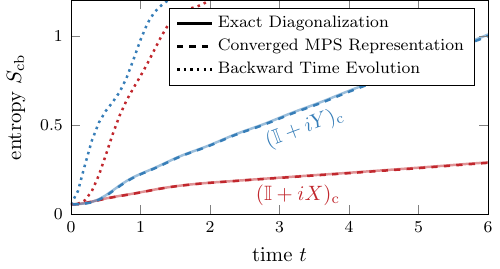}
    \caption{Comparison between the central bond entropy $S_{\mathrm{cb}}$ with the plain parallel transport disentangler obtained from \gls{ed} and the iterative GMRES method for \gls{mps} (cf.~\cref{eq:VectorizedLyapunov}) at $N=26$ sites of the TFI chain (see \cref{eq:model_heisenberg}), showing perfect agreement between the two, obtained already after one sweep in \cref{eq:VectorizedLyapunov}. While the entanglement under parallel transport for $(\idmat + iY)_{\mathrm{c}}$ grows faster than for $(\idmat + iX)_{\mathrm{c}}$, it is much less than obtainable through backward time evolution.}
    \label{fig:mps_exact_plot}
\end{figure}

\paragraph*{Parallel transport for \gls{mps}. }
To realize parallel transport in the \gls{mps} framework, we notice that the Lyapunov equation, \cref{eq:lyapunov}, can be vectorized rather naturally in the \gls{mps} framework.
Based on this observation, we introduce a Krylov-subspace~\cite{krylov1931numerical} based approach to \cref{eq:LyapunovNumerical}, by solving the vectorized matrix equation
\begin{align}    
    \begin{array}{rccclcccc}
        (                                                                                                      & \tilde A \otimes \idmat & + & \idmat \otimes \tilde A & ) & \vect \tilde G & = & - & \vect \tilde Q \\
        \left( \vphantom{\begin{tikzpicture}[baseline=(C.base)]
                                 \useasboundingbox (-0.525,-1.2-0.275) rectangle (0.525,1.2+0.275);
                                 \node (C) at (0,0) {$\vphantom{\tau}$};
                             \end{tikzpicture}} \right.                                    &
        \begin{tikzpicture}[baseline=(C.base)]
            \useasboundingbox (-0.525,-1.2-0.275) rectangle (0.525,1.2+0.275);
            \node (C) at (0,0) {$\vphantom{\tau}$};
            \node[draw,fill=lightgray!20,circle, minimum height=0.55cm,minimum width=0.55cm] (X1) at (0,1.2) {};
            \node[draw,fill=lightgray!20,circle, minimum height=0.55cm,minimum width=0.55cm] (X3) at (0,-0.4) {};
            \draw[densely dotted] (X1) -- (X3);
            \draw (X1) -- ++(-0.525,0);
            \draw (X3) -- ++(-0.525,0);
            \draw (X1) -- ++(+0.525,0);
            \draw (X3) -- ++(+0.525,0);
            \draw[dashed] (X1) ++(-0.525,-0.8) -- ++(2*0.525,0);
            \draw[dashed] (X3) ++(-0.525,-0.8) -- ++(2*0.525,0);
        \end{tikzpicture} & +                       & 
        \begin{tikzpicture}[baseline=(C.base)]
            \useasboundingbox (-0.525,-1.2-0.275) rectangle (0.525,1.2+0.275);
            \node (C) at (0,0) {$\vphantom{\tau}$};
            \node[draw,fill=lightgray!20,circle, minimum height=0.55cm,minimum width=0.55cm] (X1) at (0,0.4) {};
            \node[draw,fill=lightgray!20,circle, minimum height=0.55cm,minimum width=0.55cm] (X3) at (0,-1.2) {};
            \draw[densely dotted] (X1) -- (X3);
            \draw (X1) -- ++(-0.525,0);
            \draw (X3) -- ++(-0.525,0);
            \draw (X1) -- ++(+0.525,0);
            \draw (X3) -- ++(+0.525,0);
            \draw[dashed] (X1) ++(-0.525,+0.8) --  ++(2*0.525,0);
            \draw[dashed] (X3) ++(-0.525,+0.8) -- ++(2*0.525,0);
        \end{tikzpicture} & 
        \left.\vphantom{\begin{tikzpicture}[baseline=(C.base)]
                                \useasboundingbox (-0.525,-1.2-0.275) rectangle (0.525,1.2+0.275);
                                \node (C) at (0,0) {$\vphantom{\tau}$};
                            \end{tikzpicture}} \right)                                    & 
        \begin{tikzpicture}[baseline=(C.base)]
            \node (C) at (0,0) {$\vphantom{\tau}$};
            \node[draw,fill=lightgray!20,regular polygon, regular polygon sides=6, minimum height=0.6cm] (X1) at (0,1.2) {};
            \node[draw,fill=lightgray!20,regular polygon, regular polygon sides=6, minimum height=0.6cm] (X3) at (0,-0.4) {};
            \draw[densely dotted] (X1) -- (X3);
            \draw (X1) -- ++(-0.525,0);
            \draw (X3) -- ++(-0.525,0);
            \draw[rounded corners] (X1) -- ++(0.525, 0) -- ++(0, -0.8) -- ++(-2*0.525,0);
            \draw[rounded corners] (X3) -- ++(0.525, 0) -- ++(0, -0.8) -- ++(-2*0.525,0);
        \end{tikzpicture}
                                                                                                               & =                       & - & 
        \begin{tikzpicture}[baseline=(C.base)]
            \node (C) at (0,0) {$\vphantom{\tau}$};
            \node[draw,fill=lightgray!20,rectangle, rounded corners, minimum width=0.55cm, minimum height=0.55cm] (X1) at (0,1.2) {};
            \node[draw,fill=lightgray!20,rectangle, rounded corners, minimum width=0.55cm, minimum height=0.55cm] (X2) at (0,-0.4) {};
            \draw[densely dotted] (X1) -- (X2);
            \draw (X1) -- ++(-0.525,0);
            \draw (X2) -- ++(-0.525,0);
            \draw[rounded corners] (X1) -- ++(0.525, 0) -- ++(0, -0.8) -- ++(-2*0.525,0);
            \draw[rounded corners] (X2) -- ++(0.525, 0) -- ++(0, -0.8) -- ++(-2*0.525,0);
        \end{tikzpicture}
    \end{array},\label{eq:VectorizedLyapunov}
\end{align}
where the different shapes (circle, hexagon and rectangle) correspond to the tensors of the respective \gls{mpo} ($\tilde A$, $\tilde G$ and $\tilde Q$) with solid lines denoting their legs, dotted lines their links and dashed lines being the identity.
The equation is solved using the iterative Krylov method GMRES~\cite{Saad1986} applied to the \gls{mps}~\footnote{The numerical cost of this procedure can be thought of as very similar to the cost of a DMRG setting with the ``Hamiltonian'' $\tilde A \otimes \idmat + \idmat \otimes \tilde A$ and ``state''  $\vect \tilde G$. Here, the bond dimension of the ``Hamiltonian'' scales as $\order{d^2}$ with $d$ the bond dimension of the initial thermal purification $w$. We found $d \lesssim 30$ in practice and restricting $\chi(\vect {\tilde G}) \lesssim 30$ to be sufficient. Therefore, the cost of solving \cref{eq:VectorizedLyapunov} is practically negligible compared to the subsequent cost of real-time evolving the purification.}, which is known in the DMRG context from the correction vector method~\cite{PhysRevB.60.335,Nocera_2016}. Our practical implementation is based on the \gls{tn} library ITensor~\cite{Fishman_2022,ITensor-r0.3}. 
The resulting auxiliary Hamiltonian $\tilde{G}$ obtained from \cref{eq:VectorizedLyapunov} reproduces the exact results, as is shown in \cref{fig:mps_exact_plot}, thus demonstrating how parallel transport can be implemented efficiently in the \gls{mps} framework. The numerical data on a total system size of $N=140$ shown in \cref{fig:plots-tfi_comparison_plot} then exemplifies how full parallel transport disentanglers can be harnessed in larger scale \gls{mps} simulations. There, the geometric disentangler comes at a similar overhead cost to the simple backward time evolution approach, while at least matching the performance in entanglement mitigation of the inherently more costly \Renyi disentangler. More specifically, the main advantage of the parallel transport approach is that $\tilde G$ may be computed once for a given initial condition and physical Hamiltonian $H$, and then be used for various simulations. By contrast, the \Renyi disentangler involves solving an optimization problem at each time-step of every simulation. If this ``compute once use many times" character of $\tilde G$ is not applicable in some computational scenario, e.g. for a time-dependent physical Hamiltonian, complementary to our present approach of solving a Lyapunov equation, parallel transport may also be implemented on the fly based on singular value decomposition in every time step as outlined in \cref{sec:parallel_transport}.

A complementary approach to computing $\tilde G$ given by direct numerical evaluation of \cref{eq:aux_ham_formal} in the \gls{mps} framework is discussed in \cref{sec:integral_form_discussion}.
However, we found the above Krylov-subspace approach to the Lyapunov equation to be drastically superior, since it is both numerically more stable and requires significantly less computational resources due to its fast convergence.

\begin{figure}
    {
        \vbox to 0pt {
                \raggedright
                \textcolor{white}{
                    \subfloatlabel[1][fig:renyi_schmidt:a]
                    \subfloatlabel[2][fig:renyi_schmidt:b]
                    \subfloatlabel[3][fig:renyi_schmidt:c]
                }
            }
    }
    \includegraphics{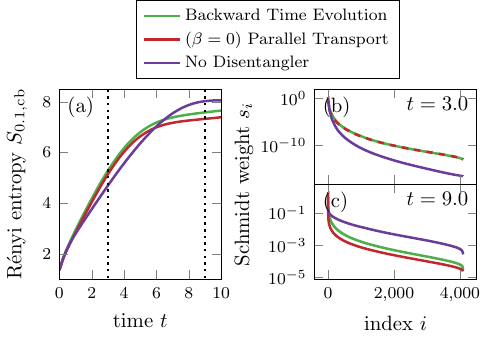}
    \caption{In \protect\subref{fig:renyi_schmidt:a} the time resolved \Renyi entropy $S_{\alpha=0.1}$ across the central bond for $N=26$ sites of the TFI chain (see \cref{eq:model_heisenberg}) obtained from \gls{ed}, showing that the tail of the Schmidt weights obtained from the infinite temperature is not significantly smaller than that of comparable methods. The corresponding Schmidt weight distribution at times $t=3.0$ in \protect\subref{fig:renyi_schmidt:b} and $t=9.0$ in \protect\subref{fig:renyi_schmidt:c} corroborates this, as for numerically relevant times \protect\subref{fig:renyi_schmidt:b} the state transported without disentangler has the fastest decaying Schmidt tail.}
    \label{fig:renyi_schmidt}
\end{figure}

\paragraph*{Infinite temperature approximation. }
From \cref{fig:beta_comparison}, approximating $\tilde G$ by its infinite temperature value (see \cref{eq:g_inf}) even at finite temperatures may be a promising and straightforward way to utilize parallel transport in existing \gls{mps} code.
This amounts to applying the conjugate operator on the mirrored site in the auxiliary degrees of freedom and evolving the state using the system Hamiltonian. The simplicity of this approximation is appealing, as virtually no restrictions on the time-evolution algorithm are imposed and therefore gate-based algorithms such as \gls{tebd} remain viable. 
While the infinite temperature approximation leads to a substantially reduced entanglement entropy (cf.~\cref{fig:entanglement_landscape:a,fig:entanglement_landscape:b}), this reduced entanglement entropy does not always reflect itself in a significantly decreased growth of the maximum or average bond dimension $\chi_{\mathrm{max}}$ and $\overline{\chi}$ when compared to the program of backward time evolution (cf. \cref{fig:entanglement_landscape:c,fig:entanglement_landscape:d}).
The local \Renyi optimization, despite leading to a similarly low von Neumann entropy as parallel transport, exhibits both significantly higher maximum and average bond dimension.
Overall, the infinite temperature approximation to parallel transport here provides a simple and viable scheme to obtain an efficient presentation of lowly entangled purified states.
However, it is fair to say that its outperformance of the backward time evolution in some scenarios may not be sufficient to compensate existing further optimization schemes of the latter~\cite{Barthel2013}.

\begin{figure*}[tb]
    \centering
    \includegraphics{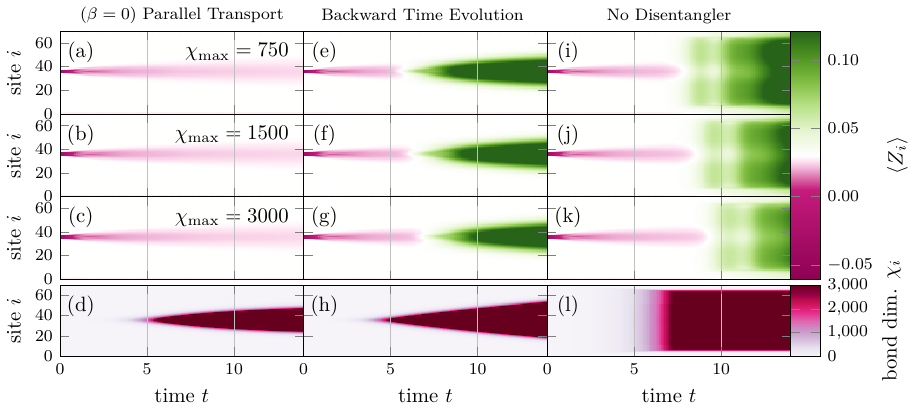}
    \caption{Expectation value of local spin $\langle Z_i\rangle$ in (a) - (c), (e) - (g), (i) - (k) and bond dimension in (d), (h), (l) after locally quenching a system with $N = 140$ spins governed by the double field Ising Hamiltonian in \cref{eq:tranverse_parallel_ising} equilibrated at $\beta = 0.1$. While the bond dimension grows light-cone like for $(\beta=0)$ parallel transport (a) - (d) and  backward time evolution (e) - (h), it increases uniformly across the chain if no disentangler is used (i) - (l). Upon overflow of the maximum bond dimension $\chi_{\mathrm{max}}$, the $(\beta=0)$ parallel transport scheme continues to produce physically reasonable data, while the other two methods show a sudden loss of the background polarization. For  the growth of entanglement entropy in this quench scenario, see inset in \cref{fig:plots-tfi_comparison_plot}.}
    \label{fig:local_quench}
\end{figure*}

To obtain a deeper understanding of the growth of the maximum bond dimension, we refer to \Renyi entropies $S_\alpha$~\cite{10.1063/1.4838856} with $\alpha \ll 1$ in \cref{fig:renyi_schmidt:a}, which quantify the tail in the distribution of Schmidt weights of a state, with higher $S_\alpha$ indicating worse approximability with sharper cutoff~\cite{Schuch_2008,Verstraete_2006}.
Indeed, both the state transported using backward time evolution and infinite temperature parallel transport exhibit higher \Renyi entropies until some saturation point in time.
This is similarly reflected in the Schmidt weights $s_i$ themselves, which are obtained by diagonalizing the reduced density matrix $\rho_L = \trace_R \ketbra{\psi}$ with eigenvalues $s_i^2$.
Here $i$ refers to the index giving the $s_i$ decreasing order.
The Schmidt weight distributions are depicted in \cref{fig:renyi_schmidt:b,fig:renyi_schmidt:c}, confirming the previous findings that the state transported without disentangler has an easier approximable Schmidt spectrum for relevant transient times in \cref{fig:renyi_schmidt:b}, while the other methods yield favorable Schmidt weight distributions at late times in \cref{fig:renyi_schmidt:c}.

\paragraph*{Application to the local quench setup.}

Entanglement growth during the real-time evolution of the post-quench dynamics with respect to the model Hamiltonian in \cref{eq:tranverse_parallel_ising} is quantified in the inset of \cref{fig:plots-tfi_comparison_plot}. Clearly, the central bond von Neumann entropy $S_{\mathrm{cb}}$ grows significantly slower in time compared to other disentanglers. Remarkably, as shown in \cref{fig:local_quench} already the $(\beta = 0)$ approximation to parallel transport for the time-evolution of an operator $\langle Z_i(t)\rangle$ outperforms both backward time evolution and using no disentangler.
While the growth in the bond dimension is slightly retarded compared to backward time evolution, it provides similar light-cone growth of the maximum bond dimension required for a high-precision approximation of the purified state.
Furthermore, upon overflow of the maximum bond-dimension, the results of the geometric disentangler continue to be physically reasonable, while both backward time evolution and plain time-evolution without disentangler suffer from qualitatively unphysical artifacts, such as the quite sudden vanishing of $\langle Z_i \rangle$. 
By contrast, the geometric disentangler does fully capture the dilution of the excitation and continued equilibration expected to occur for this fully chaotic model system even for much smaller maximum bond dimensions. This data provides evidence that the very low von Neumann entropy of the geometrically disentangled purification affords a reasonable MPS approximation, even for times at which the bond dimension has grown out of bound when admitting only a very small truncated weight.

\paragraph*{Outlook on approximate skew corrections. }
In \cref{fig:corrected-entropy}, we have shown that including skew corrections (cf. \cref{eq:entropy-linear-system}) to parallel transport may lead to a significant further reduction of entanglement. 
However, exactly including skew corrections in \gls{mps} simulations of large systems to us seems computationally unviable. 
To address this issue, we now outline how skew corrections might be used at an approximate level, which may define an interesting direction of future work. 
Inspired by variational cluster approximations (VCA)~\cite{RevModPhys.68.13,potthoff2003variational}, we propose to solve exactly for skew corrections on small clusters of auxiliary sites (the data in \cref{fig:corrected-entropy} corresponds to $N_A = 6$  auxiliary sites) at every time step, and combine the corrections from different time steps to a ``brick-wall" of correcting unitaries along the lines of the \Renyi disentangler approach \cite{PhysRevB.98.235163}. 
An advantage to VCA  methods is that imperfections of this local approximation do not directly lead to incorrect physical results but only limit the performance of the disentangling gauge transformation in mitigating the computational cost from dynamical entanglement growth. 
This advantage is shared by the local (two-site) approximation of the \Renyi disentangler, of course. 
Yet, while the latter approach solves an optimization problem brute force at every time-step, skew corrections make use of geometric insights that lead to a better scaling with cluster size than a global optimization approach (see discussion in \cref{subsec:exact_studies}). It seems tempting to incorporate skew-corrections not only with respect to the von Neumann entropy, but for arbitrary \Renyi entropies. In \cref{app:extending-skew-corrections-to-renyi-entropies} we investigate this possibility, finding evidence that this program is numerically stable in our present framework only for \Renyi entropies down to $\alpha \approx 1$.

\subsection{On the Computability of Correlation Functions}
\label{subsec:correlation_functions}

\begin{figure}
    \includegraphics{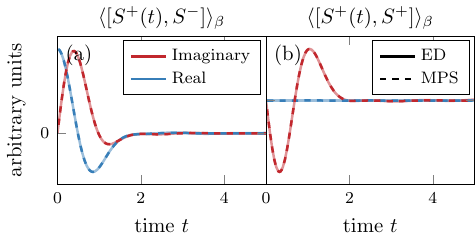}
    \caption{In (a) and (b) the components of the correlation function $\langle [S^+(t), S^-]\rangle_\beta$ and $\langle [S^+(t),S^+]\rangle_\beta$ with $S^\pm = X \pm i Y$ obtained from full \gls{ed} and the \gls{mps} method using the scheme in \cref{eq:corr_fun_retarded} (cf.\ \cref{app:spin-spin-correlation-from-unitaries}) showing perfect agreement between the exact and approximate results for cutoff $\epsilon_{\mathrm{trunc}}= 10^{-9}$ at $N=26$ sites of the TFI chain (see \cref{eq:model_heisenberg}).}
    \label{fig:correlationfn_convergence}
\end{figure}

Having investigated the intricate relationship between entanglement growth and parallel transport, we return to the objective of computing retarded correlation functions \cref{eq:thermal_expval}.
Specifically, we exemplify how spin-spin correlation functions involving the non-invertible operators $S^+,S^-$ may be obtained in our geometric disentangler framework (cf. \cref{app:spin-spin-correlation-from-unitaries} for details). As a benchmark, in \cref{fig:correlationfn_convergence}, we compare the retarded correlation function obtained from \gls{mps} methods to that obtained from full \gls{ed}, i.e. by representing the whole density matrix and explicitly computing the commutator in \cref{eq:thermal_expval}. This indicates that our assumption of invertible operators required for parallel-transport does not provide a major issue for practical calculations. 

\begin{figure}
    \includegraphics{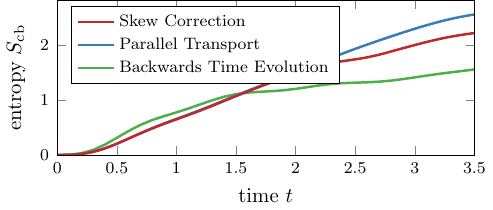}
    \caption{
        The entropy $S_{\mathrm{cb}}$ across the central bond at $N=12$ sites of the TFI chain (see \cref{eq:model_heisenberg}) for various disentanglers upon perturbation by $(\idmat + X)_{\mathrm{c}}$. 
        For this operator, skew corrections continue to improve upon plain parallel transport. 
        Despite initially mitigating entanglement growth compared to backward time evolution, the latter provides a less entangled state at later times.
    }
    \label{fig:corrected-entropy-real}
\end{figure}

The situation is more delicate when computing correlation functions that involve anticommutator contributions (cf. \cref{eq:corr_fun_advanced}).
While parallel transport mitigates growth of the von Neumann entropy after a unitary perturbation such as $(\idmat + i X)_{\mathrm{c}}$ introduced to compute the commutator contribution in \cref{eq:corr_fun_retarded}, the entropy growth upon perturbation by $(\idmat + X)_{\mathrm{c}}$~\footnote{The operator $(\idmat + \mu X)_{\mathrm{c}}$ becomes singular for $\mu = 1$. However, as the entanglement behavior is $\mu$-dependent while being sufficiently well-behaved for $\mu \rightarrow 1$, we employ $\mu =1$ for comparability with the case $\mu = i$. See \cref{app:utilizing-the-mixing-parameter-for-computations} for an in-depth discussion on the $\mu$ dependence.} in \cref{eq:corr_fun_advanced} required to compute the anticommutator contributions remains pervasive both under parallel transport and when incorporating skew-corrections (cf. \cref{fig:corrected-entropy-real}). There, backward time evolution at later times provides lower entangled states, notwithstanding their anyway substantial von Neumann entropy. In \cref{app:utilizing-the-mixing-parameter-for-computations}, we provide details on the universality of this behavior for $\abs{\mu} \neq 1$ in \cref{eq:corr_fun_advanced,eq:corr_fun_retarded}.

\section{Concluding discussion}
\label{sec:conclusions}
We have demonstrated the far reaching potential of geometric disentanglers toward taming the dynamical entanglement growth in purified quantum many-body states. To this end, we have first presented a general analysis on the relation between the differential geometry of the Hilbert-Schmidt bundle of purifications and disentangling gauge transformations for extending the scope of classical simulations on the dynamics of mixed quantum states. Based on this theoretical backbone, we have analyzed the practical performance of geometric disentanglers for computing dynamical (real time/frequency) equilibrium response functions of a paradigmatic non-integrable spin chain model with both \gls{ed} and \gls{mps} simulations.

In summary, our geometric approach may be applied at three levels of sophistication. First and most simply, at high temperatures the infinite temperature limit of Uhlmann parallel transport may be used as a disentangler, the generator of which can be expressed exactly without causing computational overhead. Second, for time-independent physical Hamiltonians, at any finite temperature the time-independent generator of Uhlmann parallel transport can be efficiently computed as an \gls{mpo} by solving a vectorized Lyapunov equation, and then be used as a geometric disentangler for real-time evolution. Third, at finite temperature, our theoretical analysis indicates that plain parallel transport, while mitigating entanglement growth, generically does not provide a locally optimal disentangler. Instead, higher order skew corrections to parallel transport are shown with \gls{ed} simulations to significantly reduce entanglement further, at least at times where finite size effects become important. However, these skew corrections in our understanding cannot be efficiently implemented in \gls{mps} simulations at a numerically exact level. To address this issue, we have outlined how cluster approximations to skew corrections may allow to tap at least some of their additional potential in future \gls{tn} algorithms.

The efficiency of the geometric disentanglers introduced in this work crucially relies on applying a unitary perturbation to a thermal initial condition. The computation of generic retarded susceptibilities involving commutators of physical operators is found to be reducible to such unitary perturbations. By contrast, correlation functions involving anti-commutators of physical operators to our present knowledge largely elude the efficient applicability of geometric disentanglers. To assess the practical added value in the field of computational physics of our general results, we have compared the performance of geometric disentanglers to other well established methods aiming at mitigating entanglement growth in purifications. 
Within the realm of unitary perturbations and commutator-based correlation functions, our approach is at least competitive with previously proposed disentanglers in reducing entanglement growth, and is found to outperform its direct alternatives in a range of settings at late times and high temperatures; however, in certain scenarios existing optimization schemes so far unique to backward time evolution~\cite{Barthel2013} may yield lower effective bond-dimensions.
Generally speaking, disentanglers can make the biggest difference for a warm to hot physical state that is very far from a pure state, for which the gauge-ambiguity of purification and with that the entire disentangling issue perishes. Furthermore, accessing late times represents the fundamental bottleneck since the residual linear entanglement growth, however small its velocity, eventually builds up an exponential wall.

\begin{acknowledgments}
    We acknowledge discussions with Sebastian Diehl and David Luitz as well as as financial support from the German Research Foundation (DFG) through the Collaborative Research Centre SFB 1143 (Project-ID 247310070), the Cluster of Excellence ct.qmat (Project-ID 390858490). Our numerical calculations have been performed at the Center for Information Services and High Performance Computing (ZIH) at TU Dresden.
\end{acknowledgments}

\appendix

\begin{figure}
    {
        \vbox to 0pt {
                \raggedright
                \textcolor{white}{
                    \subfloatlabel[1][fig:large-mps-plot:a]
                    \subfloatlabel[2][fig:large-mps-plot:b]
                    \subfloatlabel[3][fig:large-mps-plot:c]
                }
            }
    }
    \includegraphics{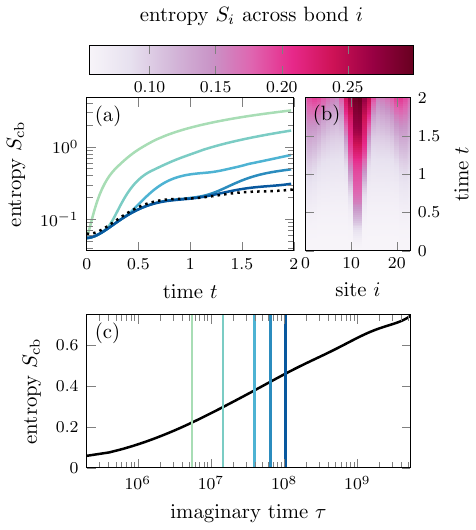}
    \caption{Slow convergence behavior of the entanglement entropy $S_{\mathrm{cb}}$ of the TFI chain (see \cref{eq:model_heisenberg}) under time evolution with the parallel transport Hamiltonian $\tilde G$ obtained from the integral \cref{eq:auxham_tensor} at $N=50$ sites in the \gls{mps} framework. In \protect\subref{fig:large-mps-plot:a} the bottlenecking entanglement entropy across the center cut is shown to decrease with larger maximum integration time (at times shown as vertical lines in \protect\subref{fig:large-mps-plot:c}) ultimately leading to the entanglement landscape in \protect\subref{fig:large-mps-plot:b}. The entanglement entropy of the \gls{mps} $T(\tau)$ (cf. \cref{eq:ttensor}) is shown in \protect\subref{fig:large-mps-plot:c}.}
    \label{fig:large-mps-plot}
\end{figure}

\section{Numerical Time-Stepping Approach to Parallel Transport}
\label{sec:parallel_transport}

A numerical on the fly approach to solving \cref{eq:parallel_transport} is obtained along the following lines~\cite{Huang_2014,Budich_2015}.
Consider the path $\gamma: t \mapsto w(t)$ at times $t$ and $t + \epsilon$ for small $\epsilon$.
Then \cref{eq:parallel_transport} turns into the finite difference equation
\begin{align}
    w^\dagger(t + \epsilon) w(t) - w^\dagger(t) w(t+\epsilon) = 0.
\end{align}
For $X = w^\dagger(t + \epsilon) w(t) = U^\dagger(t + \epsilon) \sqrt{\rho(t + \epsilon)} \sqrt{\rho(t)} U(t)$, parallel transport requires $X = X^\dagger$.
This can be used numerically to construct the unitaries $U(t + \epsilon)$ satisfying the parallel transport by inserting the singular value decomposition $\sqrt{\rho(t + \epsilon)} \sqrt{\rho(t)} = L \Sigma R^\dagger$ with $L,R^\dagger$ unitary and $\Sigma$ diagonal.
This yields
\begin{align}
    X = U^\dagger(t + \epsilon) L \Sigma R^\dagger U(t),
\end{align}
which is Hermitian if $U^\dagger(t + \epsilon) L = U^\dagger(t) R$, or alternatively if the accumulated phase $U(t + \epsilon)$ satisfies
\begin{align}
    U(t + \epsilon) = L R^\dagger U(t).
\end{align}

\section{The Integral Form of the Disentangling Hamiltonian}
\label{sec:integral_form_discussion}

A complementary approach to computing the parallel transport generator $\tilde G$ is based on the numerical evaluation of the integral in \cref{eq:auxham_w}, which amounts to contracting and integrating the tensor network
\begin{align}
    \tilde{G} = -2 \int_0^\infty \mathrm{d} \tau \;
    \begin{tikzpicture}[baseline=(C.base)]
        \node (C) at (0,0) {$\vphantom{\tau}$};
        \node[draw,fill=lightgray!20,regular polygon, regular polygon sides=6, minimum height=0.6cm] (Ul) at (0,0.4) {};
        \node[draw,fill=lightgray!20,regular polygon, regular polygon sides=6, minimum height=0.6cm] (Ll) at (0,-0.4) {};
        \node[draw,fill=lightgray!20,rectangle, rounded corners, minimum width=0.55cm,minimum height=0.55cm] (Um) at (0.8,0.4) {};
        \node[draw,fill=lightgray!20,rectangle, rounded corners, minimum width=0.55cm,minimum height=0.55cm] (Lm) at (0.8,-0.4) {};
        \node[draw,fill=lightgray!20,regular polygon, regular polygon sides=6, minimum width=0.6cm] (Ur) at (1.6,0.4) {};
        \node[draw,fill=lightgray!20,regular polygon, regular polygon sides=6, minimum width=0.6cm] (Lr) at (1.6,-0.4) {};
        \draw[densely dotted] (Ul) -- (Ll);
        \draw[densely dotted] (Ur) -- (Lr);
        \draw[densely dotted] (Um) -- (Lm);
        \draw (Ul) -- (Um) -- (Ur);
        \draw (Ll) -- (Lm) -- (Lr);
        \draw (Ul) -- ++(-0.525,0);
        \draw (Ll) -- ++(-0.525,0);
        \draw (Ur) -- ++(0.525,0);
        \draw (Lr) -- ++(0.525,0);
        \node[above, anchor=base] at (0, 0.8) {$T(\tau)$};
        \node[above, anchor=base] at (0.8,0.8) {$H$};
        \node[above, anchor=base] at (1.6,0.8) {$T(\tau)^\dagger$};
    \end{tikzpicture}
    \label{eq:auxham_tensor}
\end{align}
with the $T(\tau)$ tensor
\begin{align}
    \begin{tikzpicture}[baseline=(C.base)]
        \node (C) at (0,0) {$\vphantom{\tau}$};
        \node[draw,fill=lightgray!20,regular polygon, regular polygon sides=6, minimum width=0.6cm] (Ul) at (0,0.4) {};
        \node[draw,fill=lightgray!20,regular polygon, regular polygon sides=6, minimum width=0.6cm] (Ml) at (0,-0.4) {};
        \node[above=0.5,anchor=base] at (Ul) {$T(\tau)$};
        \draw[densely dotted] (Ul) -- (Ml);
        \draw (Ul) -- ++(-0.525,0);
        \draw (Ml) -- ++(-0.525,0);
        \draw (Ul) -- ++(+0.525,0);
        \draw (Ml) -- ++(+0.525,0);
    \end{tikzpicture}
    \equiv \exp\left(-\tau\;
    \begin{tikzpicture}[baseline=(C.base)]
            \node (C) at (0,0) {$\vphantom{\tau}$};
            \useasboundingbox (-0.525,-0.675) rectangle (1.325,0.675);
            \node[draw,fill=lightgray!20,circle,minimum width=0.55cm] (Ul) at (0,0.4) {};
            \node[draw,fill=lightgray!20,circle,minimum width=0.55cm] (Ml) at (0,-0.4) {};
            \node[draw,fill=lightgray!20,circle,minimum width=0.55cm] (Ur) at (0.8,0.4) {};
            \node[draw,fill=lightgray!20,circle,minimum width=0.55cm] (Mr) at (0.8,-0.4) {};
            \node[above=0.5,anchor=base] at (Ul) {$w$};
            \node[above=0.5,anchor=base] at (Ur) {$w^\dagger$};
            \draw[densely dotted] (Ul) -- (Ml);
            \draw[densely dotted] (Ur) -- (Mr);
            \draw (Ul) -- (Ur);
            \draw (Ml) -- (Mr);
            \draw (Ul) -- ++(-0.525,0);
            \draw (Ml) -- ++(-0.525,0);
            \draw (Ur) -- ++(0.525,0);
            \draw (Mr) -- ++(0.525,0);
        \end{tikzpicture}\right)
    \begin{tikzpicture}[baseline=(C.base)]
        \node (C) at (0,0) {$\vphantom{\tau}$};
        \node[draw,fill=lightgray!20,circle,minimum width=0.55cm] (Ul) at (0,0.4) {};
        \node[draw,fill=lightgray!20,circle,minimum width=0.55cm] (Ml) at (0,-0.4) {};
        \node[above=0.5,anchor=base] at (Ul) {$w$};
        \draw[densely dotted] (Ul) -- (Ml);
        \draw (Ul) -- ++(-0.525,0);
        \draw (Ml) -- ++(-0.525,0);
        \draw (Ul) -- ++(+0.525,0);
        \draw (Ml) -- ++(+0.525,0);
    \end{tikzpicture}\,.\label{eq:ttensor}
\end{align}
Computing this tensor network effectively amounts to performing an imaginary time evolution on $w$ with the pseudo Hamiltonian $w w^\dagger$, which can be realized using \gls{tdvp}~\cite{PhysRevLett.107.070601}.
The integral in \cref{eq:auxham_tensor} can be accomplished numerically by the midpoint rule.
While the computation of $T(\tau)$ involves only an imaginary time-evolution and is in principle efficiently doable ~\cite{barthel2017onedimensional}, the integral converges very slowly due to the flat spectrum of $w w^\dagger$ as shown in \cref{fig:large-mps-plot}. Yet, \cref{fig:large-mps-plot} nicely illustrates how the parallel transport disentangler becomes competitive with the \Renyi disentangler on better and better convergence, in agreement with \cref{fig:plots-tfi_comparison_plot}. 

\section{Correlation Functions from Phase Matching}
\label{app:correlation_functions_with_phase_matching}

Building on \cite{PhysRevB.98.235163} we approach \cref{eq:thermal_expval} with the help of two new correlation functions
\begin{align}
    C_{\pm}(\epsilon, t) & \equiv \trace [X(t)\rho_{\pm}]
\end{align}
with $\rho_- = e^{-i\epsilon Y} \rho e^{i\epsilon Y^\dagger}$ and $\rho_+ = e^{\epsilon Y} \rho e^{\epsilon Y^\dagger}$
such that the correlation function $C(t) = \langle X(t) Y \rangle$ is obtained from
\begin{align}
    C(t) = \frac{1}{2} \partial_\epsilon (C_+(\epsilon, t) + i C_-(\epsilon,t)) |_{\epsilon = 0}.
\end{align}
Now, since $\rho_\pm$ are still Hermitian matrices of full rank, their purification $w_\pm$ is well defined up to the aforementioned gauge degree of freedom $U_A$, and
\begin{align}
    C_{\pm}(t) = \trace[w_\pm^\dag X(t) w_\pm] = \trace[U_A^\dag(t) w_\pm^\dag X(t) w_\pm U_A(t)]
\end{align}
are gauge invariant functions due to the cyclic invariance of the trace. 
Hence, gauge-optimizing the time-dependent purifications $\tilde w_\pm(t) = U(t) w_\pm U_A(t)$ represents a well defined problem that can be directly addressed using the geometric toolbox of parallel transport. 
This construction conveniently solves the phase matching problem discussed in Ref.~\cite{PhysRevB.98.235163}, since the same $U_A(t)$ enters $C_{\pm}$ via both $\tilde w_\pm$ and $\tilde w_\pm^\dag$.
Numerically the exponential operators are ill-behaved, however, as they require very small cut-offs $\epsilon$ causing the accumulation of numerical dirt.

\section{Extending Skew-Corrections to \Renyi Entropies}
\label{app:extending-skew-corrections-to-renyi-entropies}

\begin{figure}
    \includegraphics{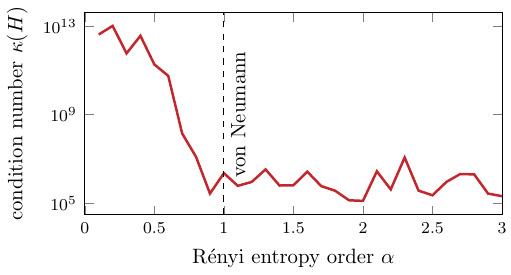}
    \caption{The condition number $\kappa(H)$ of the Hessian $H_{ij} = \partial_i \partial_j S_\alpha|_{w_0}$ for different orders of the \Renyi entropy $\alpha$ for system sizes $N=12$. The system governed by the TFI Hamiltonian (see \cref{eq:model_heisenberg}) is prepared at inverse temperature $\beta=0.2$and perturbed by $(\idmat + iX)_{\mathrm{c}}$. For the orders $\alpha$ relevant to improving \gls{mps} computations, the condition number of the problem is orders of magnitudes higher than for the accessible domain $\alpha > 1$.}
    \label{fig:conditionnumber}
\end{figure}

The considerations about skew-corrections in \cref{eq:entropy-linear-system} are independent of the chosen entropy function and as such can be adapted to include the \Renyi entropies with $\alpha < 1$ relevant to \gls{mps} methods.
One would then have to perform an initial minimization of the purification $w_0$, i.e. by gradient descent, followed by successive integration of the equation of motion \cref{eq:entropy-linear-system}.
The success of this program heavily depends on the numerical feasibility of the equations.
To quantify this, we compute the condition number $\kappa(H) = \frac{\sigma_{\mathrm{max}}(H)}{\sigma_{\mathrm{min}}(H)}$~\cite{trefethen2022numerical} of the matrix $H$ with $\sigma_{\mathrm{max}}(H)$ ($\sigma_{\mathrm{min}}(H)$) its largest (smallest) singular value.
As matrix $H$ we consider the Hessian $H_{ij} = \partial_i \partial_j S_\alpha|_{w_0}$ of the \Renyi entropies with respect to the generating parameters of the unitary transformation on the auxiliary, i.e. the second term in \cref{eq:entropy-linear-system}.
Not only is this matrix crucial in solving the equation of motion, but it is also inherent to the complexity of solving the minimization problem.
Comparing with \cref{fig:conditionnumber}, the condition number increases substantially for $\alpha < 1$, rendering the respective problems infeasible to solve numerically.

\section{Spin-spin Correlations from Unitaries}
\label{app:spin-spin-correlation-from-unitaries}

In this paper, we presented results for the spin-spin correlation functions 
\begin{align}
    C^{\epsilon_1\epsilon_2}(t,\beta) = \langle S^{\epsilon_1}(t) S^{\epsilon_2} \rangle_\beta
\end{align}
with $\epsilon_1, \epsilon_2 = \pm 1$ corresponding to the creation (annihilation) operator for $+1$ ($-1$).
The spin creation and annihilation operators $S^\pm$ are not unitary and therefore do not admit to the $(\beta=0)$ approximation, applicable to the prescription in \cref{eq:correlator_prescription}.
By writing them in terms of $X$ and $Y$ as $S^\pm = X \pm i Y$ however, the problem can be transformed into one involving only unitary operators, as 
\begin{align}
    \langle S^{\epsilon_1}(t) S^{\epsilon_2} \rangle & = \langle (X(t) + i \epsilon_1 Y(t)) (X + i \epsilon_2 Y)\rangle              \nonumber    \\
                                                     & = \langle X(t) X \rangle + i \epsilon_1 \langle X(t) Y \rangle                             \\
                                                     & \quad+ \epsilon_2 i \langle Y(t) X \rangle - \epsilon_1\epsilon_2 \langle Y(t) Y \rangle,
\end{align}
allowing for a general discussion of both full parallel transport and its approximation.

\section{Utilizing the Mixing Parameter for Computations}
\label{app:utilizing-the-mixing-parameter-for-computations}

\begin{figure}
    {
        \vbox to 0pt {
                \raggedright
                \textcolor{white}{
                    \subfloatlabel[1][fig:ed_mu:a]
                    \subfloatlabel[2][fig:ed_mu:b]
                }
            }
    }
    \includegraphics{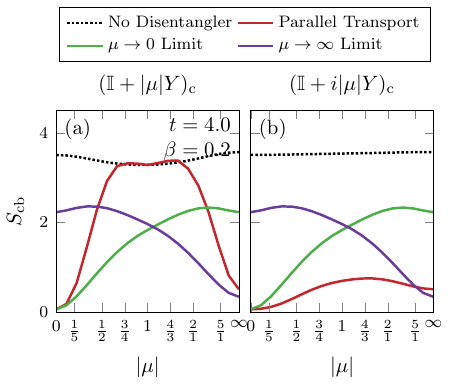}
    \caption{In \protect\subref{fig:ed_mu:a} and \protect\subref{fig:ed_mu:b} the central bond entropy $S_{\mathrm{cb}}$ for states of the TFI chain (see \cref{eq:model_heisenberg}) with $N=20$ sites at inverse temperature $\beta=0.2$ perturbed by $(\idmat + \mu Y)_{\mathrm{c}}$ with $\mu$ purely real (imaginary) in \protect\subref{fig:ed_mu:a} (\protect\subref{fig:ed_mu:b}) with $Y$ the corresponding spin operator transported using various disentanglers at time $t=4.0$ obtained from \gls{ed}. While for $\mu$ real and $\mu \approx 1$ parallel transport is as effective in mitigating the growth of $S_{\mathrm{cb}}$ as performing no disentangling, for $\abs{\mu} \rightarrow 0$ and $\abs{\mu} \rightarrow \infty$ it yields significantly decreased entanglement. For $\mu$ imaginary the entanglement observed is uniformly small. The $\mu\rightarrow 0$ and $\mu \rightarrow \infty$ approximations yield similarly low entangled states in their respective regimes.}
    \label{fig:ed_mu}
\end{figure}

In the main text of this paper we investigated parallel transport applied to the scheme in \cref{eq:hatCmu} with $\abs{\mu} = 1$, i.e. $\mu = i$ and $\mu = 1$ in \cref{eq:corr_fun_retarded,eq:corr_fun_advanced}.
This choice is arbitrary, as the scheme is analytically exact for any $\mu$.
Hence, this freedom may be chosen in numerical computations to create computationally easier accessible scenarios.
For $\rho_\mu = (\idmat + \mu Y) \rho (\idmat + \mu Y)^\dagger$ in \cref{eq:hatCmu} two limits emerge in the case that $Y$ is unitary, as it is for the operators relevant in this paper.
For $\mu \rightarrow 0$ the density matrix $\rho_\mu$ is the unperturbed $\rho$ while for $\mu \rightarrow \infty$ the density matrix $\rho_\mu$ is $Y\rho Y^\dagger$, i.e. $\rho$ but rotated by the unitary $Y$.
In the former case, backward time evolution exactly annihilates entanglement growth.
In the latter, the infinite temperature approximation of parallel transport can yield lower entangled states compared with plain parallel transport (cf. \cref{fig:beta_comparison}).
We will refer to the disentangling schemes derived from these limits as the $\mu \rightarrow 0$ and $\mu \rightarrow \infty$ disentangler respectively.

\begin{figure}
    {
        \vbox to 0pt {
                \raggedright
                \textcolor{white}{
                    \subfloatlabel[1][fig:ed_mu_err:a]
                    \subfloatlabel[2][fig:ed_mu_err:b]
                    \subfloatlabel[3][fig:ed_mu_err:c]
                    \subfloatlabel[4][fig:ed_mu_err:d]
                }
            }
    }
    \includegraphics{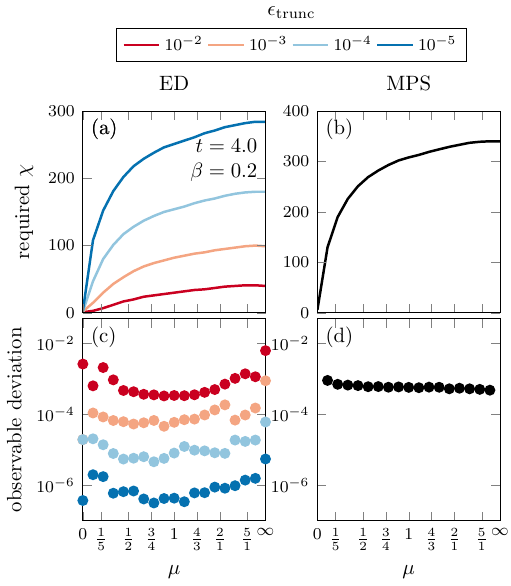}
    \caption{For thermal states of the TFI chain (see \cref{eq:model_heisenberg}) at $\beta=0.2$, $N=20$
        perturbed by $(\idmat + \mu Y)_{\mathrm{c}}$ with $\mu \in \Reals$ and transported using the $\mu \rightarrow 0$ limit to $t=4.0$ (cf. \cref{fig:ed_mu}) in \protect\subref{fig:ed_mu_err:a} the required states $\chi$ according to the truncation procedure in \cref{eq:truncation_procedure} for various truncation errors $\epsilon_{\mathrm{trunc}}$ from \gls{ed} and in \protect\subref{fig:ed_mu_err:b} the equivalent maximum bond dimension $\chi$ from \gls{mps} simulations. Both quantities show descreased numerical effort for $\delta \rightarrow 0$, while the the numerical error in the computation of the correlation function $\langle\{X(t),Y\}\rangle$ as measured by the $L^2$ distance from the exact result plateaus for fixed truncation error $\epsilon_{\mathrm{trunc}}$.}
    \label{fig:ed_mu_err}
\end{figure}

In \cref{fig:ed_mu:a,fig:ed_mu:b} we compare the entanglement accumulated by purifications transported for different parameters $\mu$.
While for $\mu$ purely imaginary (cf. \cref{fig:ed_mu:b}), the parallel transport scheme provides lowly entangled states, as observed throughout this paper, for $\mu$ purely real (cf. \cref{fig:ed_mu:a}) the entanglement growth is similar to those of states transported without disentangler.
The two limits for $\mu \rightarrow 0$ and $\mu \rightarrow \infty$ are almost indifferent to $\mu$ being purely real or purely imaginary.
As the observed entanglement growth in their domains with $\mu \approx 0$ and $\mu \approx \infty$ respectively is again very low, they may constitute a viable approach to compute the correlation function according to \cref{eq:corr_fun_advanced,eq:corr_fun_retarded}.

\begin{figure}
    \centering
    \includegraphics{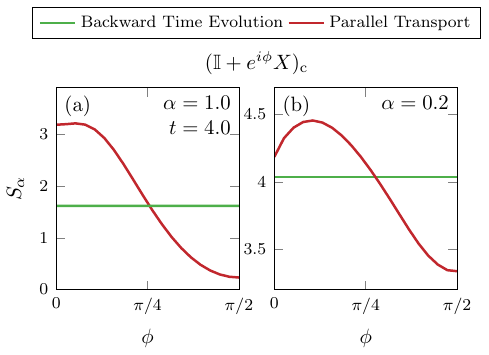}
    \caption{\Renyi{} entropies for the same setup as in \cref{fig:ed_mu}, but with the phase $\phi$ of the mixing parameter $\mu$ being varied. While the parallel transport scheme for mixing parameters close to the unitary case $\phi = \frac{\pi}{2}$ is vastly superior to backward time evolution in mitigating growth of both von Neumann entropy and the $\alpha=0.2$ \Renyi{} entropy, for arguments $\phi \lesssim \frac{\pi}{4}$, this breaks down.}
    \label{fig:eq_mu_angle}
\end{figure}

For this approach to be viable for \gls{mps}, the decreased entanglement needs to be reflected in easier approximability of the respective purified states, while retaining enough information to construct the correlation function.
This intricacy becomes the most apparent, when $\abs{\mu} = 0$ or $\abs{\mu} = \infty$, in which case the correlation function can not be recovered numerically.
Hence, we require this approach to (i) yield stable results with varying $\mu$ at decreased precision, while (ii) reducing the required bond dimension to represent the time evolved state.
In order to provide robust numerical results, we employ \gls{ed} and write the time evolved states $\ket{\psi(t)}$ in Schmidt decomposition across the central bond of the chain as 
\begin{align}
    \ket{\psi(t)} = \sum_{i=1}^N s_i \ket{l_i} \ket{r_i}
\end{align}
with $s_i$ the Schmidt weights and $\ket{l_i}$ ($\ket{r_i}$) the basis states to the left (right) of the central bond.
We truncate these states to $\ket{\tilde\psi(t)} = \sum_{i=1}^\chi s_i \ket{l_i} \ket{r_i}$ with precision $\epsilon_{\mathrm{trunc}}$ such that for some $\chi$
\begin{align}
    \sum_{i=\chi+1}^N s_i^2 < \epsilon_{\mathrm{trunc}},
    \label{eq:truncation_procedure}
\end{align}
hence mimicking the truncation procedure in \gls{mps}.
We obtain a practically relevant error of this procedure by computing correlation functions according to \cref{eq:corr_fun_retarded,eq:corr_fun_advanced} with respect to the truncated state $\ket{\tilde\psi}$ and compute their $L^2$ norm with respect to the same correlation function obtained from full \gls{ed}, i.e. by explicitly representing the density matrix $\rho$ and computing its time evolution.   

The results of this procedure are presented in \cref{fig:ed_mu_err},
indicating that (i) varying $\mu$ does not significantly increase the error of the correlation function (cf. \cref{fig:ed_mu_err:c}), while (ii) it allows to discard a lot more information from the state, indicated by much smaller $\chi$ (cf. \cref{fig:ed_mu_err:a}).
Implementing this procedure in \gls{mps} yields similar results (cf. \cref{fig:ed_mu_err:b,fig:ed_mu_err:d}).

\paragraph*{Using variable phase $\mu$ to compute the anticommutator contribution.}

Varying the phase of $\mu$ in \cref{eq:hatCmu} allows to compute the anticommutator contribution as 
\begin{align}
    C_\mu + C_{\mu^*} &= 2 \langle X(t) \rangle + 2 \abs{\mu}^2 \langle Y X(t) Y\rangle \nonumber \\
    &\qquad+ \mathrm{Re}(\mu) \langle \{ X(t), Y\}\rangle,
    \label{eq:commutatorfromphase}
\end{align}
with the all terms being computable in the parallel transport scheme.
However, comparing with \cref{fig:eq_mu_angle}, upon varying $\phi \equiv \mathrm{arg}\, \mu$, the entanglement growth increases rapidly.
This renders this approach difficult, because small phases $\phi$ correspond to small $\mathrm{Re}(\mu)$ in \cref{eq:commutatorfromphase}, therefore requiring high precision and hence invoking high numerical sensitivity.


\addcontentsline{toc}{chapter}{Bibliography}

%

\end{document}